\documentclass[12pt,a4paper]{article}
\pdfoutput=1
\usepackage{heck}
\usepackage{graphicx} 
\usepackage{cite}
\usepackage{empheq}
\usepackage{hyperref}
\usepackage{amsmath,amssymb}
\usepackage{tikz}
\usepackage{array}
\usetikzlibrary{shapes.geometric, arrows}

\tikzstyle{decision} = [
ellipse, draw, fill=blue!20, minimum height=3em, 
    text width=7em, text badly centered, node distance=3cm, inner sep=0pt]
\tikzstyle{block} = [rectangle, draw, fill=blue!20, 
    text width=5em, text centered, rounded corners, minimum height=4em]
    \tikzstyle{ir_block} = [rectangle, draw, fill=green!20, 
    text width=5em, text centered, rounded corners, minimum height=4em]
    \tikzstyle{r_block} = [rectangle, draw, fill=yellow!20, 
    text width=5em, text centered, rounded corners, minimum height=4em]
\tikzstyle{line} = [draw, -latex']
\tikzstyle{cloud} = [draw, ellipse,fill=red!20, node distance=3cm,
    minimum height=2em]

\newcommand{\beqn}{\begin{equation}\begin{aligned}}
\newcommand{\eeqn}{\end{aligned}\end{equation}}

\newcommand{\fl}{\mathfrak{l}}
\newcommand{\fn}{\mathfrak{n}}

\newcommand{\bDelta}{\overline{\Delta}}

\def\frakg{\ensuremath{\mathfrak{g}}}
\def\frakh{\ensuremath{\mathfrak{h}}}

\def\frakp{\ensuremath{\mathfrak{p}}}
\def\frakU{\ensuremath{\mathfrak{U}}}

\newcommand{\scN}{\ensuremath{\mathcal{N}}}

\newcommand{\beq}{\begin{equation}\begin{aligned}}
\newcommand{\eeq}{\end{aligned}\end{equation}}

\newlength\mytemplen
\newsavebox\mytempbox
\definecolor{myblue}{rgb}{.97,.97,1}

\makeatletter
\newcommand\mybluebox{%
    \@ifnextchar[
       {\@mybluebox}%
       {\@mybluebox[0pt]}}

\def\@mybluebox[#1]{%
    \@ifnextchar[
       {\@@mybluebox[#1]}%
       {\@@mybluebox[#1][0pt]}}

\def\@@mybluebox[#1][#2]#3{
    \sbox\mytempbox{#3}%
    \mytemplen\ht\mytempbox
    \advance\mytemplen #1\relax
    \ht\mytempbox\mytemplen
    \mytemplen\dp\mytempbox
    \advance\mytemplen #2\relax
    \dp\mytempbox\mytemplen
    \colorbox{myblue}{\hspace{1em}\usebox{\mytempbox}\hspace{1em}}}
\makeatother

\def\beq{\begin{eqnarray}}\def\eeq{\end{eqnarray}}
\def\be{\begin{equation}}\def\ee{\end{equation}}

\def\r{\rho}

\def\a{\alpha}
\def\e{\varepsilon}

\def\b{\beta}

\def\D{\Delta}

\def\l{\lambda}

\begin{document}

\preprint{IPMU17-0063}


\title{Polology of Superconformal Blocks}

\authors{\centerline{Kallol Sen and  Masahito Yamazaki}}

\date{October, 2018}

\institution{IPMU}{\centerline{Kavli IPMU, University of Tokyo, Kashiwa, Chiba 277-8583, Japan}}

\abstract{We systematically classify all possible poles of superconformal blocks as a function of the scaling dimension of intermediate operators, for all superconformal algebras in dimensions three and higher. This is done by working out the recently-proven irreducibility criterion for parabolic Verma modules for classical basic Lie superalgebras. The result applies to correlators for external operators of arbitrary spin, and indicates presence of infinitely many short multiplets of superconformal algebras, most of which are non-unitary. 
We find a set of poles whose positions are shifted by linear in $\scN$ for $\scN$-extended supersymmetry.
We find an interesting subtlety for 3d $\scN$-extended superconformal algebra with $\scN$ odd associated with odd non-isotropic roots. We also comment on further applications to superconformal blocks.}


\pagenumbering{Alph}
\maketitle 
\pagenumbering{arabic}
\tableofcontents

\section{Introduction}

The long-distance physics of supersymmetric field theories is generically described by 
superconformal field theories (SCFTs). In this paper we study superconformal blocks,
which capture the universal features of correlation functions of
SCFTs. The superconformal blocks are also crucial inputs to the 
to the conformal bootstrap program \cite{Ferrara:1973yt,Polyakov:1974gs} and its recent reincarnation \cite{Rattazzi:2008pe, Rychkov:2009ij}, as applied to SCFTs.\footnote{The literature is too vast to be exhaustive here. A sample of early references motivated by the works of \cite{Rattazzi:2008pe, Rychkov:2009ij} include  \cite{Rattazzi:2010gj,Poland:2010wg,Poland:2011ey,Beem:2013qxa,Bashkirov:2013vya,Fitzpatrick:2014oza,Berkooz:2014yda,Khandker:2014mpa,Alday:2014qfa,Chester:2014fya,Li:2014gpa,Beem:2014zpa,Bobev:2015jxa}, see also \cite{Liendo:2016ymz,Lemos:2016xke} for more recent computations of superconformal blocks.}

Since superconformal blocks is determined purely from superconformal algebra (SCA), 
it is natural to ask if we can directly derive the superconformal block from the representation theory of SCAs,
thereby establishing a bridge between the physics of SCFTs and the mathematics of representation theory of SCAs.
In this paper we make a first step in this direction, and we systematically 
list all the (possible) poles of the conformal blocks,
as a function of a scaling dimension $\Delta$ of the intermediate operator.

The basic logic behind this is the follows \cite{Penedones:2015aga,Yamazaki:2016vqi}.\footnote{The two-dimensional counterpart of this argument, as applied to the Virasoro algebra, goes back to the classic paper by Zamolodchikov \cite{Zamolodchikov:1985ie}.}
The conformal block describes the four-point function of operator $\mathcal{O}_{1}, \ldots, \mathcal{O}_4$,
which four-point function can be expressed as
\begin{align}
\begin{split}
&\langle \mathcal{O}_1(x_1) \mathcal{O}_2(x_2) \mathcal{O}_3(x_3) \mathcal{O}_4(x_4) \rangle \\
&=\sum_{\alpha=\mathcal{O}, P^{\mu}\mathcal{O}, \ldots ,  \, ; \, \mathcal{O}\in \mathcal{O}_1 \times \mathcal{O}_2}
\frac{\langle \mathcal{O}_1(x_1) \mathcal{O}_2(x_2) | \alpha \rangle
 \langle \alpha | \mathcal{O}_3(x_3) \mathcal{O}_4(x_4) \rangle
 }{\langle \alpha | \alpha \rangle}   \ , 
 \label{alpha_decomp}
 \end{split}
\end{align}
where the sum is now over all the superconformal descendants $\alpha$
of the superconformal primary  $\mathcal{O}$.

Since the three-point functions never diverge as a function of the scaling dimension $\Delta$, the 
divergence can arise only if one of the descendants $\alpha$ of the superconformal primary $\mathcal{O}$ becomes null. This is exactly when the representation of the SCA spanned by descendants, which is known mathematically as the parabolic Verma module, is reducible.\footnote{This fact is the basis for the derivation of the recursion relation \cite{Kos:2013tga,Kos:2014bka,Penedones:2015aga,Iliesiu:2015akf,Costa:2016xah} for conformal blocks in \cite{Penedones:2015aga}.}

The problem is then to work out precisely when the parabolic Verma module is reducible (or irreducible).
We can solve this problem by analyzing the recently-proven irreducibility criterion of \cite{Oshima:2016gqy}.
The result is summarized in Section \ref{sec.D}.

The rest of this paper is organized as follows. In Section \ref{sec.SCA} we summarize some aspects of representation theories of SCAs, 
and in particular introduce irreducibility criterion of \cite{Oshima:2016gqy}.
In Section \ref{sec.irreducibility} we derive from the irreducibility criterion
systematic algorithms to identity the locations of poles of superconformal blocks.
We then apply the algorithm to SCFTs in Sections \ref{sec.3} and \ref{sec.D}. We conclude with some remarks on
unitarity bounds (Section \ref{sec.unitarity}) and superconformal blocks (Section \ref{sec.remarks}).  We include appendices on root systems (Appendices \ref{app.ABD} and \ref{sec.SCA}).


Let us also include brief comments specifically for more mathematics-oriented readership.
Note that while our motivation comes partly from physics, our results can be stated mathematically, without referring to physics. 
Let $\mathfrak{g}$ be a SCA, which we take to be one of the SCAs (see Appendix \ref{app.SCA}).
We the consider the parabolic Verma module $M_{\mathfrak{p}}(\lambda)$ with highest weight
associated with a parabolic subalgebra $\mathfrak{p}$ (see \cite{Oshima:2016gqy} for definitions).
Then the question is to identify the values of the highest weight $\lambda$ such that the 
resulting representation is reducible. This is a natural supersymmetrization of the 
setup of  \cite{EHW}, who studied unitary irreducible highest weight representations
of parabolic Verma modules associated with Hermitian symmetric spaces.
We will spell out the irreducibility criterion of \cite{Oshima:2016gqy} explicitly for all the SCA examples,
and determine all the highest weights where the parabolic Verma module is reducible.\footnote{We do not work out the decompositions into irreducible components.
This is related with the question of computing the Kazhdan-Lusztig polynomials \cite{KazhdanLusztig} 
for parabolic Verma modules, which seem to be unknown for the cases at hand, at least in general.}

\section{Superconformal Algebra}\label{sec.SCA}

In this section we briefly summarize the basics of SCAs and their parabolic Verma modules relevant to SCFTs. 
We follow the notations of \cite{Oshima:2016gqy}.
Readers familiar with SCAs can safely skip Section \ref{subsec.Lie}, and 
those familiar with \cite{Oshima:2016gqy} can skip Sections \ref{subsec.root} and \ref{subsec.parabolic}.

\subsection{Lie Superalgebra}\label{subsec.Lie}

In this paper we study the representation theory of the 
SCA $\mathfrak{g}$, in spacetime dimensions greater than two.
The superconformal algebras $\mathfrak{g}$ is a Lie superalgebra,\footnote{See e.g.\ \cite{KacAdvM,Frappat:1996pb,Musson} for mathematical introduction to Lie superalgebras.}
whose even/bosonic (odd/fermionic) degree part we denote by 
$\mathfrak{g}_0$ ($\mathfrak{g}_1$). 
SCA  has been classified by Nahm \cite{Nahm:1977tg} long ago.
The SCA exists only in $D\le 6$, and are given 
as one of the following Lie superalgebras\footnote{We here list the complex form of the Lie superalgebra.}, of type A, B, D or F:\footnote{In this paper, we choose the convention that the argument of $\mathfrak{sp}$ is even, e.g.\ $\mathfrak{sp}(2)\simeq \mathfrak{su}(2)$.}\footnote{Since the irreducibility criterion of \cite{Oshima:2016gqy} applies in general to any contragredient finite-dimensional Lie superalgebra with an indecomposable Cartan matrix, it is straightforward to repeat the computations in this paper for 
those other such Lie superalgebras which do not appear in the list of SCAs, 
for example for $C(n) = \mathfrak{osp}(2| 2n-2)\, (n\ge 2)$,  $D(2,1;\alpha)\,  (\alpha \ne 0, 1)$ and $G(3)$. Note that some of these symmetries do appear when we consider 
defects in SCFTs, so that we break some of the Poincar\`e symmetries. For example, a $1/2$-BPS Wilson loop for 5d $\mathcal{N}=1$ SCFT discussed in \cite{Assel:2012nf} preserve a subgroup $D(2,1;2)\oplus \mathfrak{su}(2)$ of $F(4)$.}
\begin{align}
\begin{split}
&A(m,n) = \mathfrak{sl}(m+1| n+1) \;, \quad m>n\ge 0 \;,\\
&A(n,n) =\mathfrak{psl}(n+1| n+1) \;, \quad n\ge 1 \;, \\
&B(m,n) = \mathfrak{osp}(2m+1| 2n) \;, \quad m\ge 0, n>0 \;, \\
&D(m,n) = \mathfrak{osp}(2m| 2n) \;, \quad m\ge2, n\ge 1 \;, \\
&F(4) \;. 
\end{split}
\end{align}
\begin{itemize}
\item 
The 3d SCA is 
\begin{align}
\begin{split}
&\mathfrak{g}^{\rm 3d}=\mathfrak{osp}(\scN |4)
=\begin{cases}
B\left(\frac{\scN-1}{2}, 2\right) & (\scN \textrm{: odd}) \\
D\left(\frac{\scN}{2}, 2 \right)& (\scN \textrm{: even})\\
\end{cases} \;,\\
&\mathfrak{g}^{\rm 3d}_0=\mathfrak{so}(\scN)\oplus
\mathfrak{sp}(4) \;,
\end{split}
\end{align}
with $\scN=1,2, \ldots 8$.\footnote{3d SCFT with $\scN=7$ supersymmetry automatically has $\scN=8$ supersymmetry \cite{Bashkirov:2011fr,Cordova:2016emh}.}

\item 
The 4d SCA is
\begin{align}
\begin{split}
&\mathfrak{g}^{\rm 4d}=A(3, \scN-1)
=\begin{cases}
\mathfrak{su}(4| \scN) & (\scN=1,2,3) \\
\mathfrak{psu}(4|4) &(\scN=4) \\
\end{cases} \;,\\
&\mathfrak{g}^{\rm 4d}_0=
\begin{cases}
\mathfrak{su}(4)\oplus
\mathfrak{su}(\scN-1) & (\scN=1,2,3) \\
\mathfrak{su}(4)\oplus
\mathfrak{su}(\scN-1)\ominus \mathfrak{u}(1) &(\scN=4)\\
\end{cases} \;.
\end{split}
\end{align}

\item 
The 5d $\mathcal{N}=1$ SCA is
\begin{align}
\begin{split}
&\mathfrak{g}^{\rm 5d}=F(4)\;,\\
&\mathfrak{g}^{\rm 5d}_0=\mathfrak{so}(7)\oplus
\mathfrak{sp}(2) \;.
\end{split}
\end{align}
It is known that 5d $\mathcal{N}=2$ SCA does not exist.

\item 
The 6d SCA is 
\begin{align}
\begin{split}
&\mathfrak{g}^{\rm 6d}=\mathfrak{osp}(8|2 \scN)=D(4, \scN)\;,\\
&\mathfrak{g}^{\rm 6d}_0=\mathfrak{so}(8)\oplus
\mathfrak{sp}(2\scN) \;,
\end{split}
\end{align}
with $\mathcal{N}=1,2$.\footnote{In the literature these are more often called $\scN=(1,0)$ and $\scN=(2,0)$ SCAs, to 
emphasize the chirality. $\scN=(1,1)$ case is excluded in Nahm's classification.}

\end{itemize}
In this paper we exclude the case of $D=2$, where
we have an infinite-dimensional Virasoro symmetry and its extensions.

The details of SCAs depend on spacetime dimensions and the number of supersymmetries, but in all the cases SCA is generated by dilatation $D$, rotation $J$, translation $P$, superconformal transformation $K$, R-symmetry generator $R$, 
supersymmetry $Q$, superconformal $S$, with $\mathfrak{g}_0$ generated by $D, J, P, K, R$ and 
$\mathfrak{g}_1$ by $Q, S$. Their commutation relations are summarized for example in \cite{Minwalla:1997ka}.
Schematically and neglecting numerical coefficients and index structures, the commutation relation takes the form
\begin{align}
\begin{split}
&[D,P]\sim P\;,\quad 
[D,K]\sim K\;,\quad
[D,Q]\sim Q\;,\quad
[D,S]\sim S\;,\\
&[J,P]\sim P\;,\quad
[J,J]\sim J\;,\quad
[J,Q]\sim Q\;,\quad
[J,S]\sim S\;,\\
&[P,S]\sim Q\;,\quad
[K,Q]\sim S\;,\quad
\{Q, Q \}\sim P\;,\quad
\{S, S \}\sim K\;,\quad
\{Q, S \}\sim D+J+R\;.\quad
\end{split}
\end{align}

In this paper, when quoting the mathematical results 
we will refer to the compact form of the conformal algebra
$\mathfrak{g}=\mathfrak{so}(D+2)$, rather than the actual conformal algebra in the Lorentzian signature
$\mathfrak{so}(D, 2)$; the same applies to SCAs, 
and for example for 3d SCA we will refer to $\mathfrak{g}=\mathfrak{osp}(\scN |4)$
rather than $\mathfrak{g}=\mathfrak{osp}(\scN |2,2)$. 
However, this is only for convenience and we will discuss the Lorentzian superconformal algebra $\mathfrak{so}(D, 2)$. Recall that $\mathfrak{so}(D, 2)$ can equally thought of as the Euclidean conformal algebra
$\mathfrak{so}(D+1,1)$ in the radial quantization
\begin{align}
P^{\dagger}=K, \quad Q^{\dagger}=S, \quad J^{\dagger}=J, \quad D^{\dagger}=-D \;,
\label{radial}
\end{align} 
so that some of the generators of SCAs are not Hermitian \cite{Minwalla:1997ka}.

\subsection{Root System}\label{subsec.root}

For our considerations a crucial concept is the root system of a Lie superalgebra.

Let us denote the set of roots by $\Delta$. We then have the 
root space decomposition 
\beq
\frakg = \frakh \oplus \bigoplus_{\alpha\in \Delta}
\frakg^{\alpha} \;,
\eeq
where $\frakh$ be a Cartan subalgebra of $\frakg_0$, and $ \frakg^{\alpha}$ is the root space corresponding to $\alpha$.

The set of roots $\Delta$ has three different decompositions.

The first decomposition is into and even/bosonic (odd/fermionic) roots $\Delta_{0}$ ($\Delta_{1}$),
mirroring the decomposition of $\mathfrak{g}$ into $\mathfrak{g}_0$ and $\mathfrak{g}_1$:
\beq
\Delta= \Delta_0 \cup \Delta_1\;.
\label{decomp.01}
\eeq

The second decomposition is into positive and negative roots:
\beq
\Delta = \Delta^{+} \cup \Delta^{-}\;.
\label{delta_pm}
\eeq
In practice, the choice of the positive roots of determined by an
ordering of simple roots (see the examples in the later sections).

To describe the third decomposition, 
let us define $\overline{\Delta}_0$ and 
the set of isotropic roots $\overline{\Delta}_1$ by
\begin{align}
\overline{\Delta}_0 :=\{
\alpha \in \Delta_0 |\, \alpha/2 \not \in \Delta_1
\}\subset \Delta_0 \;,\quad
\overline{\Delta}_1 :=\{
\alpha \in \Delta_1 |\, 2\alpha \not \in \Delta_0
\} \subset \Delta_1\;.
\label{overline_def}
\end{align}
It is known that an odd root $\alpha$ is isotopic if and only if $(\alpha, \alpha)=0$.
The set of non-isotropic roots is then defined to be the complement of $\bDelta_1$
\be
\Delta_{\rm non-iso}:= \Delta_0 \, \cup \, (\Delta_1 \setminus \overline{\Delta}_1) \;,
\ee
so that 
\begin{align}
\Delta=\Delta_{\rm non-iso}\cup \overline{\Delta}_1 \;.
\label{decomp.iso}
\end{align}
This is the third decomposition, which will be crucial for the discussion of irreducibility criterion below.

For most SCAs
we have $\Delta_{1} =  \overline{\Delta}_{1}$ and $\Delta_{\rm non-iso}= \Delta_0$,
so that the two decompositions \eqref{decomp.01} and \eqref{decomp.iso} coincide.
The only exceptions are the cases of 3d $\mathcal{N}$-extended SCA
with $\scN$ odd ($\mathfrak{g}=\mathfrak{osp}(\scN|4)$).
In particular, for $\scN=1$ case we have
$\overline{\Delta}_{1} =  \varnothing$
and $\Delta_{\rm non-iso}= \Delta_0 \, \cup \, \Delta_1=\Delta$.

This distinction between isotropic odd roots and non-isotropic odd roots are of crucial importance in the analysis below, since 
the two types of odd roots appear differently in the irreducibility criterion of \cite{Oshima:2016gqy}, as will be explained below.
While the difference between isotropic odd roots and non-isotropic odd roots are well-known in mathematical literature (and in fact 
such a distinction is one of the crucial new structures in the representation theory of Lie superalgebras as opposed to that of Lie algebras), 
this subtlety seems to have been missed in the more physics-oriented literature. In this paper we perform a complete analysis taking this point into account.

\subsection{Parabolic Verma Module}\label{subsec.parabolic}

In SCFT correlation functions, a superconformal primary running in the intermediate channel has a 
specific scaling dimension, spin(s) and R-charge(s). Equivalently, a superconformal primary is a finite-dimensional
irreducible representation $V(\lambda)$ of
the reductive subalgebra generated by $D, J$ and $R$:
\begin{align}
\fl:=\langle J, D, R \rangle =\mathfrak{so}(D)\oplus \mathfrak{so}(2)\oplus \mathfrak{g}_R\subset \mathfrak{g}_0 \;,
\label{Levi}
\end{align}
where $\mathfrak{g}_R$ is the R-symmetry algebra of the theory and 
$\lambda$ is a  highest weight of the representation, and collectively denotes the scaling dimension, spin(s) and R-charge(s).
This subalgebra $\fl$ is called the Levi subalgebra, and can also be written as
\beq
\fl  = \frakh\oplus \bigoplus_{\alpha\in \Delta_{\fl}} \frakg^{\alpha}, 
\eeq
where $\Delta_{\fl}$ is a subset of even simple roots of $\mathfrak{g}$ corresponding to generators of $\fl$.

Superconformal primaries are by definition annihilated by generators $K_{\mu}$ and $S_{\alpha}$\footnote{Annihilation by $K_{\mu}$ and $S_{\alpha}$ holds only at the origin.} ,
which generate a subalgebra $\fn$:
\beq
\fn :=  \bigoplus_{\alpha\in \Delta_{\fn}} \frakg^{-\alpha}=\langle K, S \rangle \;,
\eeq
where we defined\footnote{Logically this might better be denoted $\Delta_{\fn}^{+}$, however we in this paper do not use
$\Delta_{\fn}^{+}$ without the plus sign and hence we simply dropped the plus sign, to simplify the notation.}
\begin{align}
\Delta_{\fn}:=\Delta^+\setminus \Delta_{\fl} \;.
\end{align}

We can naturally combing the two ingredients above.
The representation $V(\lambda)$ of $\fl$ can be 
lifted to representation of a parabolic subalgebra $\mathfrak{p}$
\begin{align}
\mathfrak{p}
:= \frakh\oplus \bigoplus_{\alpha\in \Delta_{\fl} \cup \Delta^+} \frakg^{\alpha}
= \fl \oplus \fn \;, 
\end{align}
by letting $\mathfrak{n}$ act trivially on $V(\lambda)$.
This representation can further be extended canonically to a representation of the 
universal enveloping algebra 
$\frakU(\mathfrak{p})$.
Then the parabolic Verma module is defined by
\begin{align}
M_{\mathfrak{p}}(\lambda):=\frakU(\mathfrak{g}) \otimes_{\frakU(\mathfrak{p}) } V({\lambda}) \;.
\end{align}
This is the representation relevant to the superconformal blocks, as emphasized in \cite{Yamazaki:2016vqi}.

\section{Irreducibility Criterion}\label{sec.irreducibility}

\subsection{First Algorithm}\label{subsec.first}

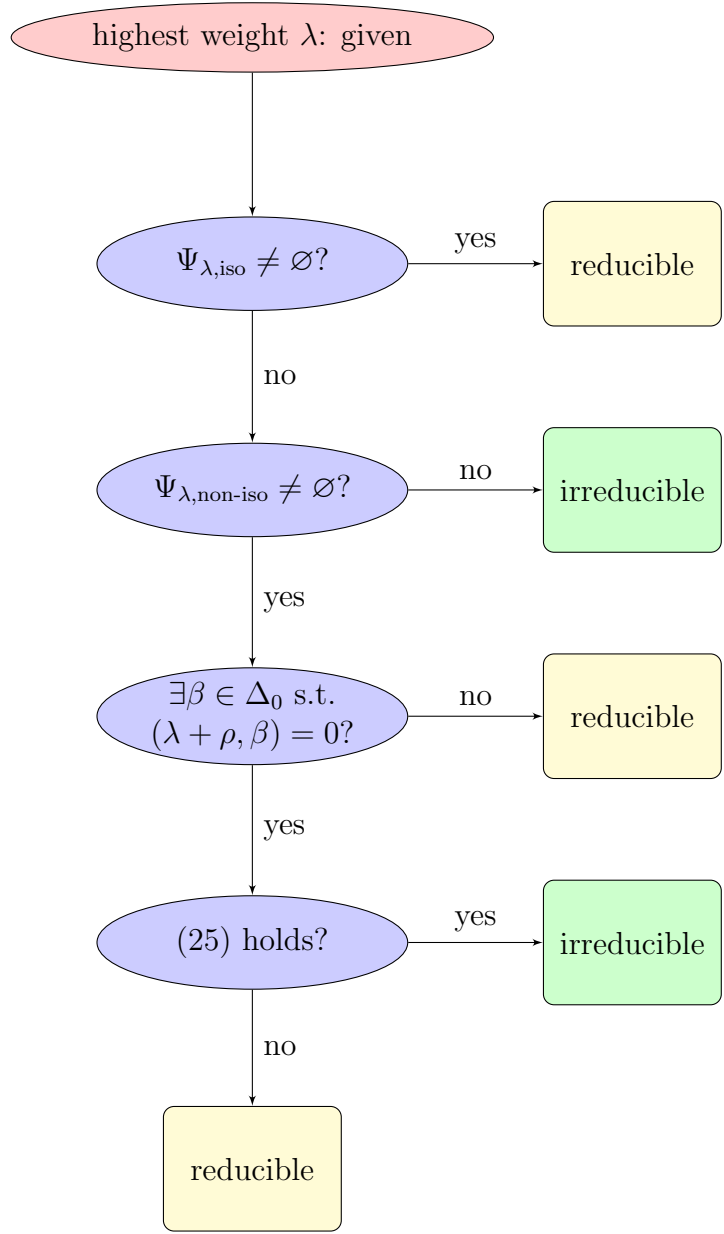
\begin{figure}
\caption{Practical algorithm for checking irreducibility/reducibility of $M_{\mathfrak{p}}(\lambda)$.
In principle we can directly go to the last step and check \eqref{eq.final}, however it is in practice much more economical to proceed in this way. Steps 2 and 3 are useful in practice, however they can safely be skipped and we can 
directly choose to proceed to Step 4, right after Step 1. This algorithm will further be simplified for the case of $\scN$-extended SCA, as we will discuss later.}
\label{thm.simple}
\label{fig.flow_chart}
\begin{center}
\begin{tikzpicture}[node distance = 2 cm, auto]
    \node [cloud] (lambda) {highest weight $\lambda$: given};
    \node [decision, below of=lambda] (iso) {$\Psi_{\lambda, \textrm{iso}}\ne \varnothing$?};
    \node [r_block, right of=iso, node distance=5cm] (r1) {reducible};
    \node [decision, below of=iso] (non-iso) {$\Psi_{\lambda, \textrm{non-iso}}\ne \varnothing$?};
    \node [ir_block, right of=non-iso, node distance=5cm] (r2) {irreducible};
    \node [decision, below of=non-iso] (Weyl) {$\exists \beta\in \Delta_0$ s.t. $(\lambda+\rho, \beta)=0$?
    };
    \node [r_block, right of=Weyl, node distance=5cm] (r3) {reducible};
    \node [decision, below of=Weyl, node distance=3cm] (final) {\eqref{eq.final} holds?};
     \node [ir_block, right of=final, node distance=5cm] (r4) {irreducible};
    \node [r_block, below of=final, node distance=3cm] (r5) {reducible};
    \path [line] (lambda) -- (iso);
    \path [line] (iso) -- node {yes}(r1);
    \path [line] (iso) -- node {no}(non-iso);
    \path [line] (non-iso) -- node {no}(r2);
    \path [line] (non-iso) -- node {yes}(Weyl);
    \path [line] (Weyl) -- node {no}(r3);
    \path [line] (Weyl) -- node {yes}(final);
     \path [line] (final) -- node {yes}(r4);
     \path [line] (final) -- node {no}(r5);
\end{tikzpicture}
\end{center}
\end{figure}

The necessary and sufficient condition for irreducibility of the parabolic Verma module 
was already derived in \cite{Oshima:2016gqy}, in particular Theorem 3 therein (this generalizes the Jantzen's irreducibility criterion of parabolic Verma modules for Lie algebras \cite{Jantzen} to Lie superalgebras\footnote{There are mathematical papers explicitly working out Jantzen criterion for scalar (spin zero) parabolic Verma modules for semisimple Lie algebras. See e.g.\ \cite{He_reducibility} for recent discussion.}).
The criterion however involves complicated linear combinations of characters of the superconformal algebra, and can be become rather messy in examples, if we try to use explicit formulas for the characters in the most straightforward manner. 
In this paper, we therefore would follow the 
systematic procedure outlined in Fig.~\ref{fig.flow_chart}.

One remark worth mentioning is that our criterion is different from the Kac's irreducibility criterion \cite{Kac_LNM}, which is often invoked in the literature of superconformal field theories. The criterion of \cite{Oshima:2016gqy} reduces to Kac's criterion \cite{Kac_LNM} when the Levi subalgebra is a Borel subalgebra, however for applications to the physics of SCFTs the Levi algebra 
should be given by \eqref{Levi}, and is not a Borel subalgebra.

\paragraph{Step 1}

Let us define the set $\Psi_{\lambda, {\rm iso}}$ by
\begin{empheq}[box={\mybluebox[7pt]}]{equation}
\Psi_{\lambda, {\rm iso}}:=
\left\{
\beta \in \overline{\Delta}_{1}^+ \Big|
\, ( \lambda+\rho, \beta )=0 
\right\}  \;,
\label{iso}
\end{empheq}
where $\rho$ is the Weyl vector defined by
\beq
\rho:=
\rho_0-\rho_1 \;, \quad
\rho_0:=\frac{1}{2} 
\sum_{\alpha \in \Delta^+_{0}} \alpha  \;, \quad
\rho_1:=\frac{1}{2} 
\sum_{\alpha \in \Delta^+_{1} }\alpha  \;,
\label{Weyl}
\eeq
and $(-, -)$ is the canonical non-degenerate $\mathfrak{g}$-invariant pairing (contravariant form) defined by BPZ conjugation in radial quantization \eqref{radial}.\footnote{As will become clear, the overall normalization factor of this pairing is irrelevant for the considerations of this paper, see e.g.\ \eqref{non-iso}.} We also used the notations such as $\overline{\Delta}_{1}^+:=\overline{\Delta}_{1}\cap \Delta^+$ and $\Delta^+_{0}:=\Delta^+ \cap \Delta_{0}$, so that more indices mean intersections of corresponding sets.

If this set $\Psi_{\lambda, {\rm iso}}$ is non-empty, then Proposition 4 of \cite{Oshima:2016gqy} guarantees that the parabolic Verma module is reducible. Otherwise we proceed to the next step.

Note that this step is essentially the Kac criterion of \cite{Kac_LNM},  and is already studied in the previous literature, see e.g.\ \cite{Dobrev:1985qv,Dobrev:1985vh,Dobrev:1985qz,Minwalla:1997ka,Dobrev:2002dt}. This is however not the end of the story, as we commented a few paragraphs above.

\paragraph{Step 2}

Let us define the set $\Psi_{\lambda, {\rm non-iso}}$ by
\begin{empheq}[box={\mybluebox[7pt]}]{equation}
\begin{split}
\Psi_{\lambda, {\rm non\mathchar`-iso}}:=
&\left\{
\beta \in \Delta_{\fn } \cap \overline{\Delta}_0 \Big|
\, n_{\alpha}:=
 \frac{2( \lambda+\rho, \alpha )}{(\alpha, \alpha)} \in \mathbb{Z}_{>0}
\right\} \\
&\bigcup
\left\{
\beta \in \Delta_{1}^+ \setminus  \overline{\Delta}_{1}^+ \Big|
\, n_{\beta}:=\frac{2( \lambda+\rho, \alpha )}{(\alpha, \alpha)}
 \in 2\mathbb{Z}_{>0}-1 \right\}  \;,
\end{split}
\label{non-iso}
\end{empheq}
If this set is empty, then Proposition 5 of \cite{Oshima:2016gqy} guarantees that the parabolic Verma module is irreducible. If the set is non-empty, we proceed to the next step.

\paragraph{Step 3}

In the third step, we check if there exists an even root $\beta\in \Delta_0$
which is orthogonal to $\lambda+\rho$: 
\be
(\lambda+\rho, \beta)=0 \;.
\label{hyperplane}
\ee
This is equivalent with the condition that $\lambda+\rho$ is on the boundary of the Weyl chambers.\footnote{Weyl group for a Lie superalgebra is defined to be the Weyl group for its even part $\mathfrak{g}_0$.}
If this condition is not satisfied, then the module is irreducible thanks to Corollary 1 of \cite{Oshima:2016gqy}; otherwise we proceed to the last step.

\paragraph{Step 4}

In this final step, we use the Theorem 3 of \cite{Oshima:2016gqy}, adopted here for the case $\Psi_{\lambda, {\rm iso}}=\varnothing$:
the parabolic Verma module $M_{\mathfrak{p}}(\lambda)$ is irreducible
if and only if
\begin{empheq}[box={\mybluebox[7pt]}]{equation}
\sum_{\alpha \in \Psi_{\lambda, {\rm non\mathchar`-iso}}}
 \chi^{\mathfrak{p}}(s_{\alpha}.\lambda )=0 \;.
\label{eq.final}
\end{empheq}
Here
\beq
\label{chi_def}
\chi^{\mathfrak{p}}(\lambda):= \sum_{w\in W_{\fl}} \textrm{det}(w)\, \textrm{ch}\, M(w. \lambda)\;, 
\eeq
where $W_{\fl}$ is the Weyl group of $\fl$, 
and the dot action $w.\lambda$ is defined to be
the Weyl group action, with shift by the Weyl vector $\rho$:
\beq
w.\lambda:=w(\lambda+\rho)-\rho \;.
\label{dot_shift}
\eeq

The condition \eqref{eq.final} is more complicated than the conditions in the previous three steps.
However, first note that the value of $\lambda$ in this last step is already much constrained by the condition in Step 3, and the set $\Psi_{\lambda, {\rm non\mathchar`-iso}}$ is typically a small set.

What is crucial moreover is the fact that we do not need to 
use explicit expressions for the characters $\textrm{ch}\, M(\mu)$, and all we need is 
that the character $\textrm{ch}\, M(\mu)$'s with different highest weights $\mu$ are linearly independent.\footnote{This is basically how Propositions 4 and 5 of \cite{Oshima:2016gqy} are derived from the irreducibility criterion (Theorem 3 in \cite{Oshima:2016gqy}.} In this respect, it is useful to know that 
\begin{align}
\chi^{\mathfrak{p}}(w.\lambda)=\det(w)\chi^{\mathfrak{p}}(\lambda)\;,
\label{chi_sign}
\end{align}
for any Weyl reflection $w\in W_\fl$ for the Levi subalgebra $\fl$,
and in particular $\chi^{\mathfrak{p}}(\lambda)=0$ if $w.\lambda=\lambda$ and $(-1)^{w}=-1$.
In fact, we can prove in general \cite{Oshima:2016gqy} that any cancellation of the form \eqref{eq.final}, if it holds, can be proven by repeated use of \eqref{chi_sign}.

Moreover, we know already from \eqref{hyperplane} that $\lambda+\rho$ is orthogonal to $\beta$, and hence $s_{\alpha} (\lambda +\rho)$
is in the hyperplane orthogonal to $s_{\alpha}(\beta)$.

Let us add that there are simplified versions of the irreducibility criteria, see~\cite[Proposition 4 and 5]{Oshima:2016gqy}. These are either sufficient but not necessary, or necessary but not sufficient conditions,
however are stated purely in the language of root systems and hence can be useful, especially when the 
rank of $W_{\fl}$ is large.

Let us also emphasize again that while we have chosen to present the algorithm in four steps, 
this is purely for practical convenience, and we can for example skip Steps 2 and 3.

\subsection{Improved Algorithm}\label{subsec.improved}

The algorithm mentioned above is very concrete, and as we will see later
can be worked out case by case (we will work some examples in the next section).
Since there are only finitely many SCAs (excluding the non-supersymmetric conformal algebras, which exists in 
an arbitrary spacetime dimensions), this will enough for our purpose.

However, when working out some examples we will notice that we end up repeating the same computations multiple times, and consequently we will realized that there is a more efficient method to implement this algorithm.
This builds on the fact that the algorithm has already been implemented and solved for the non-supersymmetric CFTs in 
general spacetime dimensions \cite{Penedones:2015aga}.

Let us here consider $\scN$-extended supersymmetry in a given spacetime dimension ($D=3,4,5,6$).
Let us hereafter exclude the case of 3d $\mathcal{N}$-extended supersymmetry with $\scN$ odd.\footnote{As we will see later, the same type of argument does apply to 3d $\mathcal{N}$-extended supersymmetry with $\scN$ odd cases, reducing the analysis to the case of 3d $\mathcal{N}=1$ SCA.
}
Then (as commented below \eqref{overline_def}) $\Delta_{\rm non-iso}=\Delta_0$ and the set $\Delta_{\fn}\cap \overline{\Delta}_0$ is given by the translation generators ($P_{\mu}$'s), and hence is independent of the value of $\mathcal{N}$. 

This does not necessarily mean that the condition in \eqref{eq.final} is independent of the value of $\mathcal{N}$.
First of all, the expression for $\lambda+\rho$ is different for different values of $\scN$.
We can denote this vector as a linear span of roots corresponding to 
dilatation $D$, spin $J_i$, and R-symmetry generators $R_a$ (recall $\Delta$ is the scaling dimension of the superconformal primary)\footnote{
There is an unfortunate crash of notation, where $\Delta$ is used both for scaling dimension and the root system (both are standard notations, in physics and mathematics, respectively). We hope that context will make clear
which we mean. Note that any $\Delta$ with indices (such as $+$ and $0$) will a subset of the root system.
}:
\begin{align}
\lambda+\rho =(-\Delta+c_D) \beta_D +\sum_i c_{J_i} \beta_{J_i}+\sum_a c_{R_a} \beta_{R_a}\;.
\label{lambdarho}
\end{align}
To be more precise, detailed form of the expressions such as in \eqref{lambdarho} 
depends on the specific spacetime dimensions and the amount of supersymmetries, but 
such distinctions will not affect the analysis below. The notation in \eqref{lambdarho} follows the 
$D=3$ case, but the analysis in other dimensions is similar.

Coming back to the main track of the discussion, depending on the values of $\mathcal{N}$, we have the following changes:
\begin{itemize}
\item[(a)] The R-symmetries are different for difference values of $\mathcal{N}$, and hence
we have different number of generators $R_a$. Their coefficients $R_a$ are also affected by the value of $\mathcal{N}$.
\item[(b)] The coefficient $c_D$ depends linearly on the value of $\mathcal{N}$.
\end{itemize}

However, there is an important point: \emph{the value of $c_{J_i}$ is independent of $\mathcal{N}$}.
Indeed, the only possible source for the change of $c_{J_i}$ comes from
the change of the Weyl vector $\rho$ \eqref{Weyl}, and in particular its odd part $\rho_1$. 
However, this is independent of $\beta_{J_i}$, since
if there is a supercharge with a certain spin there is also another supercharge with opposite spin, 
as required by the invariance under the symmetry $J_i\to -J_i$ (this is either a parity or a rotation symmetry,
depending on the parity of the spacetime dimension).

Let us now come to the consequences of this fact for the algorithm of the previous subsection.
First, the change (a) makes no difference in the analysis of Step 2 and later
steps of the algorithm, since the roots in $\Delta_{\fn}\cap \overline{\Delta}_0$ are all translation generators and hence are orthogonal to the newly-introduced R-symmetry generators. 
The effect of (b) is simply to linearly shift the values of $\Delta$, but not the values of spins, 
which we can easily take into account. It turns out that this is the {\it only} change for the analysis of \eqref{eq.final},
namely for the analysis of Steps 2, 3 and 4.

Note that the actual expression for the character $\textrm{ch}\, M(\lambda)$,
and hence of $\chi^{\frakp}(\lambda)$ (see \eqref{chi_def}), heavily depends on the value of $\mathcal{N}$.
However, as we explained around \eqref{chi_sign},
in the analysis of Step 4 we did not need the explicit form of the characters, 
but rather only the relations \eqref{chi_sign}, which is not affected since the Levi subgroup $\fl$ \eqref{Levi} (and its Weyl group $W_{\fl}$) are independent of the amount of supersymmetry.

A careful reader might have also recognized the following subtlety:  the even root $\beta$ in Step 3 
can be one of the roots for R-symmetry generators $R$, or more generally those in linear combination with
the roots for the rotations $J$. These roots are absent for $\mathcal{N}=0$ case, and hence
the analysis of Step 3 does seem to depend on the value of $\scN$.
Notice, however, that (as explained previously)
Step 3 is included for pure convenience and we can instead directly go to Step 4 without losing anything, 
and then we can easily see that the presence of the R-symmetry generators do not affect the analysis of \eqref{eq.final}, since R-symmetry generators belong to the Levi subgroup
$\mathfrak{l}$ and hence do not appear in the definition of $\Psi_{\lambda, {\rm non-iso}}$.

The bottom line of the discussion to this point is that, up to the shift of the values of the scaling dimension $\Delta$,
and modulo such a shift the Steps 2, 3, 4 now reduce to the analysis of $\mathcal{N}=0$ cases.\footnote{For tensor representations this was already worked out in \cite{Penedones:2015aga}.}
Summarizing the discussion, we arrived at the following improvement of the algorithm.

\paragraph{Step 1}

This is the same as Step 1 before.

\paragraph{Step 2$^\prime$}

This subsumes all of the Steps 2, 3 and 4 in the previous algorithm.

Start with the results for the non-supersymmetric case (see Appendix \ref{app.N0}). Compute $\lambda+\rho$ to obtain the shift of the scaling dimension, as originating from $\mathcal{N}$, and shift the result appropriately. Note we do this shift only in Step 2$^\prime$, and not in Step 1.

\section{\texorpdfstring{Example: $D=3$}{Example: D=3}}\label{sec.3}

In this section, let us first work out the case of three spacetime dimensions. In this section we provide many details,
to illustrate the ideas and the algorithms of the previous sections in concrete examples.
From the representation-theory viewpoint, this spacetime dimension is also the most interesting example,
since this include $\mathcal{N}$-extended supersymmetry with $\scN$ odd, which contains odd non-isotropic roots (i.e.\ odd roots whose length is non-zero).

For the remainder, we will adhere to the notations in appendix \ref{app.SCA}, which summarize the 
necessary data for SCAs.\footnote{In this paper we choose $\lambda$ to be the highest weight, rather than the lowest weight, to match the representation theory conventions in \cite{Oshima:2016gqy}. Some physics literature, e.g.\ \cite{Minwalla:1997ka}, rather use lowest weights. The two conventions are related by a replacement $\lambda\to -\lambda$.}

\subsection{\texorpdfstring{$D=3,\  \mathcal{N}=0$}{D=3, N=0}}\label{sec.3dN0}
Let us start with the case of $\mathcal{N}=0$. 
The analysis will be the same as in \cite{Penedones:2015aga}, except our analysis here includes spinor representations.

The root system is given by
\be
\D_0=\bigg\{\pm \b_D\,, \ \pm\b_J\,, \ \pm\b_D\pm\b_J\bigg\}\,,\quad \D_1=\varnothing \;,
\ee
with positive roots given by
\be
\D^+=\bigg\{\b_D\,, \ \b_J\,, \ \b_D\pm\b_J\bigg\}\,.
\ee
The inner product between roots is given by
\be
(\beta_a, \beta_b)=2 \delta_{a,b} \;, \quad a,b=D,J \;.
\label{3dN0_pairing}
\ee

The Levi subgroup $\mathfrak{l}$ is the $\mathfrak{so}(3)$
subgroup of $\mathfrak{so}(5)$, and thus
\begin{align}
&\D_{\mathfrak{l}}=\bigg\{\pm\b_J\bigg\}\;, \\
&\D_{\fn}\cap \overline{\D}_0=\D_{\fn}\cap \D_0=\D_{\fn}=\D^{+}\setminus \D_{\fl}=\bigg\{\b_D,\b_D\pm\b_J\bigg\}\,. 
\end{align}
The set $\D_{\fn}\cap \overline{\D}_0$ represent the three translation generators $P_3, P_{\pm}=P_1\pm i P_2$ generating descendants. 
These operators have have dimension $1$ and spin $1, 0, -1$, as represented by the coefficients in front of $\beta_D$ and $\beta_J$.

We will employ the highest weight state given by
\be
\l=-\D\b_D+ \frac{\ell}{2} \b_J \;,
\label{3dN0_lambda}
\ee
with $\Delta$ the scaling dimension and $\ell \in \mathbb{Z}_{\ge 0}$ the angular spin;
$\ell$ is odd for a spinor representation.
The Weyl vector is given by
\be
\r=\r_0=\frac{1}{2}\sum_{\a\in\D^{+}}\a=\frac{3}{2}\b_D+\frac{1}{2}\b_J\;,
\ee
and hence
\begin{align}
\l+\r=\left(-\D+\frac{3}{2}\right)\b_D+\left(\frac{\ell}{2}+\frac{1}{2}\right)\b_J\;.
\label{3dN0_rho_sum}
\end{align}
Since there is no odd root, there is nothing to do at Step 1.

\paragraph{Step 2}
Let us compute the expressions $n_{\beta}$ for three elements of $\Delta_{\fn}$:
\begin{align}\label{ni}
\begin{split}
n_{\b_D+\b_J}=\frac{\ell}{2}+2-\D\,,\quad
 n_{\b_D}=3-2\D\,,\quad
n_{\b_D-\b_J}=1-\frac{\ell}{2}-\D\,.
\end{split}
\end{align}
From this we learn that the set $\Psi_{\lambda, {\rm non-iso}}$ is non-empty in the following cases:
\be\label{redu_even}
\D=\begin{dcases}
\frac{\ell}{2} +1 , \frac{\ell}{2}, \ldots, 2\,, &\Psi_\l^+=\{\b_D+\b_J\}\,, \\
1, 0, -1, \ldots, 1-\frac{\ell}{2} \,, &\Psi_\l^+=\{\b_D+\b_J, \b_D \}\,, \\
1-\frac{\ell}{2}-n\ \ (n=1,2, \ldots)\,,&\Psi_\l^+=\{\b_D+\b_J, \b_D, \b_D-\b_J\}\,,\\
\frac{3}{2}-n\,, \ \ (n=1,2, \ldots) \,, &\Psi_\l^+=\{\b_D\}\,,\\
\end{dcases}
\ee
for $\ell$ even, and 
\be\label{redu_odd}
\D=\begin{dcases}
\frac{\ell}{2} +1 , \frac{\ell}{2}, \ldots, \frac{3}{2}\,, &\Psi_\l^+=\{\b_D+\b_J\}\,, \\
\frac{1}{2}, -\frac{1}{2},  \ldots, 1-\frac{\ell}{2} \,, &\Psi_\l^+=\{\b_D+\b_J, \b_D \}\,, \\
1-\frac{\ell}{2}-n\ \ (n=1,2, \ldots)\,,&\Psi_\l^+=\{\b_D+\b_J, \b_D, \b_D-\b_J\}\,,\\
2-n\,, \ \ (n=1,2, \ldots) \,, &\Psi_\l^+=\{\b_D\}\,,\\
\end{dcases}
\ee
for $\ell$ odd. Notice that the first line in \eqref{redu_even} is absent for the special case of $\ell=0$.

\paragraph{Step 3}

We wish to identify the cases where $\lambda+\rho$ is orthogonal to one of the even roots.
Since the coefficient of $\beta_J$ is positive, $\lambda+\rho$ is never orthogonal to $\beta_J$,
and we only need to consider the remaining positive roots $\beta_D+\beta_J, \beta_D$ and $\beta_D-\beta_J$.

When one of these three roots are orthogonal the corresponding integer $n_{\beta}$, one of the three integers in \eqref{ni}, is zero.
At the same time, we need to make sure at at least one of the two remaining two $n$'s to be positive integers, so that
$\Psi_{\lambda, {\rm non-iso}}$ to be non-empty.
This happens only when
\be
\beta=\b_D-\b_J\,,\quad \D=1-\frac{\ell}{2}\,,\quad \Psi_{\lambda, {\rm non-iso}}=\{\b_D,\b_D+\b_J\}\,.\\
\label{3dN0_step3_1}
\ee 
for $\ell$ even; for $\ell$ odd we also need to consider the case
\be
\beta=\b_D\,,\quad \D=\frac{3}{2}\,,\quad \Psi_{\lambda, {\rm non-iso}}=\{\b_D+\beta_J\}\,.\\
\label{3dN0_step3_2}
\ee 

\paragraph{Step 4}

The final task is to verify the irreducibility criterion for \eqref{3dN0_step3_1} and \eqref{3dN0_step3_2}.

For the case of \eqref{3dN0_step3_1}
$\lambda+\rho=(\ell+\frac{1}{2})(\beta_D+\beta_J)$, leading to
\begin{align}
\begin{split}
s_{\beta_J} (s_{\b_D}(\lambda+\rho))=\left(\ell+\frac{1}{2}\right)s_{\beta_J}(-\beta_D+\beta_J)
=-\left(\ell+\frac{1}{2}\right)(\beta_D+\beta_J)
=s_{\b_D+\beta_J}(\lambda+\rho) \;,
\end{split}
\label{chi_comp_1}
\end{align}
to find (recall the shift by the Weyl vector $\rho$ in the definition of the dot action in \eqref{dot_shift})
\begin{align}
\chi^{\mathfrak{p}}(s_{\b_D+\beta_J}. \lambda) 
=\chi^{\mathfrak{p}}(s_{\beta_J}. (s_{\b_D}.(\lambda+\rho))) 
=-\chi^{\mathfrak{p}}(s_{\b_D}.(\lambda+\rho))
\;.
\label{chi_comp_2}
\end{align}
where we used \eqref{chi_sign}.
Hence \eqref{eq.final} is satisfied, and the module is irreducible.

For the case of \eqref{3dN0_step3_2}, $\Psi_{\lambda, {\rm non-iso}}$
contains only a single element, and there is non cancellation in \eqref{eq.final}
and the module is reducible after all.

Our conclusion therefore is that the module is reducible at all the values \eqref{redu_even}, \eqref{redu_odd}, except at $\D=1-\ell$.
Namely 
\begin{align}
\begin{split}
&\D=\left(\bigg\{ \frac{\ell}{2}+2-\mathbb{Z}_{>0} \right\} \setminus \left\{ 1- \frac{\ell}{2} \bigg\}  \right)  
\bigcup\left\{ \frac{3}{2}- \mathbb{Z}_{>0} \right\} \;.
\label{3dN0_result_1}
\end{split}
\end{align}  
for $\ell$ even, and 
\begin{align}
\begin{split}
&\D=\left(\bigg\{ \frac{\ell}{2}+2-\mathbb{Z}_{>0} \right\} \setminus \left\{ 1- \frac{\ell}{2} \bigg\}  \right)  
\bigcup\left\{ 1- \mathbb{Z}_{>0} \right\} \;.
\label{3dN0_result_2}
\end{split}
\end{align}  
for $\ell$ odd.

\subsection{\texorpdfstring{$D=3,\ \mathcal{N}=1$}{D=3, N=1}}

Let us next consider the case $\scN=1$.

The root system is given by
\be
\D_0=\bigg\{\pm\b_D,\pm\b_J,\pm\b_D\pm\b_J\bigg\}\,, \ \ \D_1=\left\{\pm\frac{1}{2}\b_D\pm\frac{1}{2}\b_J\right\}\,.
\ee
and thus 
\be
\overline{\D}_0=\bigg\{\pm\b_D,\pm\b_J\bigg\}\,, \ \ \overline{\D}_1=\varnothing \,.
\ee
and in particular all the odd roots are non-isotropic. 
The Levi subgroup  $\fl$ is generated by rotations $\D_\mathfrak{l}=\{\pm \b_J\}$,
and hence we have $\D_\fn=\{\b_D,\b_D\pm\b_J, \frac{1}{2}\b_D\pm\frac{1}{2}\b_J\}$.

The inner product between the roots is given by \eqref{3dN0_pairing}, as before.

The positive roots are taken to be
\be
\Delta^{+}= \bigg\{\b_D, \b_J, \b_D\pm\b_J \,, \frac{1}{2}\b_D\pm\frac{1}{2}\b_J\bigg\}\,.
\ee

The sets appearing in $\Psi_{\lambda, {\rm non-iso}}$ in \eqref{non-iso}
are given by
\be
\overline{\D}_{\fn}\cap \Delta_0=\bigg\{\b_D\bigg\}, \quad
\D_1^+ \setminus \overline{\D}_1^+=\left\{\frac{1}{2}\b_D\pm\frac{1}{2}\b_J\right\}\,.
\ee
Note in particular that we have odd non-isotropic roots.

The highest weight state is given by \eqref{3dN0_lambda} as before,
while the Weyl vector given by 
\be
\r_0=\frac{3}{2}\b_D+\frac{1}{2}\b_J\,, \ \ \r_1=\frac{1}{2} \b_D\,, \ \ \r=\r_0-\r_1=\b_D+\frac{1}{2}\b_J\,.
\ee
and thus 
\be
\lambda+\r=\left(-\Delta+1 \right) \b_D+\left(\frac{\ell}{2} +\frac{1}{2}\right)\b_J\,.
\label{3dN1_rho_sum}
\ee

\paragraph{Step 2}
We compute
\begin{align}
\begin{split}
\ \ n_{\frac{1}{2}\b_D+\frac{1}{2}\b_J}=2\left(\frac{3}{2}-\D+\frac{\ell}{2}\right)\,,\quad
  \ \ n_{\b_D}=2(1-\D)\,, \quad
 \ \ n_{\frac{1}{2}\b_D-\frac{1}{2}\b_J}=2\left(\frac{1}{2}-\D-\frac{\ell}{2}\right)\,.
\end{split}
\label{ni_2}
\end{align}
Recall (see \eqref{non-iso}) that the we want to impose the condition that 
$n_{\b_D}\in \mathbb{Z}_{>0}$ and $n_{\frac{1}{2}\b_D \pm \frac{1}{2}\b_J}\in 2\mathbb{Z}_{>0}-1$.
This means $\Psi_{\lambda, {\rm non-iso}}$ is a non-empty set at the values
\begin{align}
\begin{split}
\D=\begin{dcases}
\frac{\ell}{2}+1,\frac{\ell}{2}, \frac{\ell}{2}-1 ,\ldots  , 1 \,, \ & \Psi_{\l, {\rm non-iso}}=\left\{\frac{1}{2}\b_D+\frac{1}{2}\b_J\right\}\,,\\
0, -1, \ldots, -\frac{\ell}{2}+1 \,, \ & \Psi_{\l,{\rm non-iso}}=\left\{\frac{1}{2}\b_D+\frac{1}{2}\b_J, \b_D\right\}\,,\\
-\frac{\ell}{2}+1-n, \ (n=1,2,\dots)\,, \ & \Psi_{\l,{\rm non-iso}}=\left\{\frac{1}{2}\b_D+\frac{1}{2}\b_J,\b_D, \frac{1}{2}\b_D-\frac{1}{2}\b_J\right\}\,, \\
\frac{3}{2}-n, \ (n=1,2,\dots)\,, \ & \Psi_{\l,{\rm non-iso}}=\left\{\b_D\right\}\,.
\end{dcases}
\end{split}
\end{align} 
for $\ell$ even and positive, 
\begin{align}
\begin{split}
\D=\begin{dcases}
\frac{\ell}{2}+1,\frac{\ell}{2}, \frac{\ell}{2}-1 ,\ldots  , \frac{3}{2} \,, \ & \Psi_{\l, {\rm non-iso}}=\left\{\frac{1}{2}\b_D+\frac{1}{2}\b_J\right\}\,,\\
\frac{1}{2}, -\frac{1}{2}, \ldots, -\frac{\ell}{2}+1 \,, \ & \Psi_{\l,{\rm non-iso}}=\left\{\frac{1}{2}\b_D+\frac{1}{2}\b_J, \b_D\right\}\,,\\
-\frac{\ell}{2}+1-n, \ (n=1,2,\dots)\,, \ & \Psi_{\l,{\rm non-iso}}=\left\{\frac{1}{2}\b_D+\frac{1}{2}\b_J,\b_D, \frac{1}{2}\b_D-\frac{1}{2}\b_J\right\}\,, \\
1-n, \ (n=1,2,\dots)\,, \ & \Psi_{\l,{\rm non-iso}}=\left\{\b_D\right\}\,.
\end{dcases}
\end{split}
\end{align} 
for $\ell$ odd, 
\begin{align}
\begin{split}
\D=\begin{dcases}
1,0, -1 ,\ldots  , 1 \,, \ & \Psi_{\l, {\rm non-iso}}=\left\{\frac{1}{2}\b_D+\frac{1}{2}\b_J\right\}\,,\\
1-n, \ (n=1,2,\dots)\,, \ & \Psi_{\l,{\rm non-iso}}=\left\{\frac{1}{2}\b_D+\frac{1}{2}\b_J,\b_D, \frac{1}{2}\b_D-\frac{1}{2}\b_J\right\}\,, \\
\frac{3}{2}-n, \ (n=1,2,\dots)\,, \ & \Psi_{\l,{\rm non-iso}}=\left\{\b_D\right\}\,.
\end{dcases}
\end{split}
\end{align} 
for $\ell=0$.

\paragraph{Step 3}

As in the case of 3d $\mathcal{N}=0$ case in the previous subsection, we search for the case where
one of the three integers \eqref{ni_2} is zero, while at least one of them is positive (positive and odd for $n_{\frac{1}{2}\b_D\pm \frac{1}{2}\b_J}$).
This happens for 
\begin{align}
\begin{split}
&\D=1 \,, \ \Psi_{\l, {\rm non-iso}}=\left\{\frac{1}{2}\b_D+\frac{1}{2}\b_J\right\}\,, \quad \lambda+\rho=\left(\frac{\ell}{2}+\frac{1}{2}\right)\beta_J \;,\\
&\D= -\frac{\ell}{2} +\frac{1}{2}\,,  \ \Psi_{\l, {\rm non-iso}}=\left\{\frac{1}{2}\b_D+\frac{1}{2}\b_J, \b_D\right\}\,, \quad \lambda+\rho=\left(\frac{\ell}{2}+\frac{1}{2}\right)(\beta_D+\beta_J) \;.
\end{split}
\end{align} 

\paragraph{Step 4}
We can easily verify that the module is irreducible in the two cases above, similar to the cases analyzed in the $\scN=0$ subsection.

We therefore came to the conclusion that the possible poles of the 
3d $\mathcal{N}=1$ superconformal blocks are located at
\begin{align}
\begin{split}
&\D=\left(\bigg\{ \frac{\ell}{2}+2-\mathbb{Z}_{>0} \right\} \setminus \left\{1, 1- \frac{\ell}{2} \bigg\}  \right) \bigcup \bigg\{ \frac{3}{2}-\mathbb{Z}_{>0} \bigg\} \;.
\label{3dN1_result_1}
\end{split}
\end{align}  
for $\ell$ even, and 
\begin{align}
\begin{split}
&\D=\left(\bigg\{ \frac{\ell}{2}+2-\mathbb{Z}_{>0} \right\} \setminus \left\{1, 1- \frac{\ell}{2} \bigg\}  \right) \bigcup \bigg\{ 1-\mathbb{Z}_{>0} \bigg\} \;.
\label{3dN1_result_2}
\end{split}
\end{align}  
for $\ell$ odd. This is almost the same as the $\scN=0$ result in 
\eqref{3dN0_result_1} and \eqref{3dN0_result_2}, except the module is now reducible at $\Delta=1-\ell$ and irreducible at $\Delta=1$; this difference goes away for the special case of $\ell=0$.

\subsection{\texorpdfstring{$D=3, \ \mathcal{N}\ge 2$ with $\mathcal{N}$ Even}{D=3, N>=2 with N Even}}

Let us assume that $\mathcal{N}$ is even.
The algebra for this case is $\mathfrak{g}=\mathfrak{osp}(\mathcal{N} |4)=D\left(\frac{\scN}{2},2\right)$. 
The relevant root systems for this case is summarized in appendix \ref{sec.3d_even}.
In this case, note that all the odd roots are all isotropic: $\D_1=\overline{\D}_1$. 

The highest weight vector $\l$ for this case is given by,
\begin{align}
\l=-\D \b_J+\frac{\ell}{2}\b_J+\sum_{a=1}^{\frac{\scN}{2}} \lambda_a \delta_a \,,
\label{3dN_even_lambda}
\end{align}
where $\lambda_a\in \mathbb{Z}/2$ is given in terms of the $\mathfrak{so}(\scN)$ Dynkin labels $k_a \in \mathbb{Z}$ by the relation
\eqref{lambda_delta_even}.
The Weyl vector is given by,
\begin{align}
\r=\frac{3-\mathcal{N}}{2}\b_D+\frac{1}{2}\b_J+\sum_{a=1}^{\frac{\mathcal{N}}{2}} \frac{\mathcal{N}-2a}{2}\delta_i \,.
\label{3dN_even_rho}
\end{align}
and thus
\begin{align}
\l+\r=\bigg(\frac{3-\mathcal{N}}{2}-\D\bigg)\b_D+\bigg(\frac{\ell}{2}+\frac{1}{2}\bigg)\b_J+\sum_{a=1}^{\mathcal{N}/2} \bigg(k_a+\frac{\mathcal{N}-2a}{2}\bigg)\delta_a\,.
\label{3dN_even_rho_sum}
\end{align}

\paragraph{Step 1} 
The result of this step is that the module is reducible at values \cite{Minwalla:1997ka}
\begin{align}
\Delta
=\begin{dcases}
2-a+\frac{\ell}{2}+k_a &  \left(\a=\frac{1}{2}\b_D+\frac{1}{2}\b_J-\delta_a \right) \;, \\
1-a-\frac{\ell}{2}+k_a & \left(\a=\frac{1}{2}\b_D-\frac{1}{2}\b_J-\delta_a\right)  \;, \\
2-\scN+a+\frac{\ell}{2}-k_a & \left(\a=\frac{1}{2}\b_D+\frac{1}{2}\b_J+\delta_a\right) \;, \\
1-\scN+a-\frac{\ell}{2}-k_a  &\left(\a=\frac{1}{2}\b_D-\frac{1}{2}\b_J+\delta_a\right) \;. 
\end{dcases}\label{3d_even_iso}
\end{align}

\paragraph{Step 2$^{\prime}$} 
For the consideration of the set $\Psi_{\l, {\rm non-iso}}$,
the set  $\Delta_{\fn}\cap \overline{\Delta}_0$ is the same as that for $\mathcal{N}=0$ case, as mentioned above;
this set is orthogonal to all the R-symmetry roots $\alpha_i$.
The net effect is therefore that the value of $\Delta$ is shifted by 
$\mathcal{N}/2$,
and hence the reducible points are are given by
\begin{align}
\Delta=\left(\left\{ \ell+2-\frac{\scN}{2}-\mathbb{Z}_{>0}\right\} \setminus \left\{-\ell-\frac{\scN}{2}+1 \right\}\right)\bigcup \left(\frac{3}{2}-\frac{\scN}{2}-\mathbb{Z}_{>0}\right) \;.
\label{3d_even_noniso}
\end{align}
for $\ell$ even and 
\begin{align}
\Delta=\left(\left\{ \ell+2-\frac{\scN}{2}-\mathbb{Z}_{>0}\right\} \setminus \left\{-\ell-\frac{\scN}{2}+1 \right\}\right)\bigcup \left(1-\frac{\scN}{2}-\mathbb{Z}_{>0}\right) \;.
\label{3d_odd_noniso}
\end{align}
for $\ell$ odd.

The complete set of reducible points are obtained by combining 
\eqref{3d_even_iso}, \eqref{3d_even_noniso} and \eqref{3d_odd_noniso}.

\subsection{\texorpdfstring{$D=3, \ \mathcal{N}\ge 2$ with $\mathcal{N}$ Odd}{D=3, N>=2 with N Odd}}

Let us next consider the case of general odd $\scN$.
The relevant root systems for this case is summarized in appendix \ref{sec.3d_odd}.
In this case, we have odd non-isotropic roots, $\pm \frac{\beta_D}{2} \pm \frac{1}{2}\beta_J$,
which exist for all values of $\mathcal{N}$.

The highest weight vector $\l$ and the Weyl vector are given by,
\begin{align}
&\l=-\D \b_J+ \frac{\ell}{2} \b_J+\sum_{a=1}^{\frac{\scN-1}{2}} \lambda_a \delta_a\,,\label{3dN_odd_lambda}\\
&\r=\frac{3-\scN}{2}\b_D+\frac{1}{2}\b_J+\sum_{a=1}^{\frac{\mathcal{N}-1}{2}} \frac{\mathcal{N}-2a}{2} \delta_a \,,
\label{3dN_odd_rho}
\end{align}
where the half-integers $\lambda_a$'s are given in term of the Dynkin labels as \eqref{lambda_delta_odd}.
We thus have
\be
\l+\r=\bigg(\frac{3-\mathcal{N}}{2}-\D\bigg)\b_D+\bigg(\frac{\ell}{2}+\frac{1}{2}\bigg)\b_J+\sum_{a=1}^{\frac{\mathcal{N}-1}{2}} \bigg(\lambda_a+\frac{\mathcal{N}-2a}{2}\bigg) \delta_i\,.
\label{3dN_odd_rho_sum}
\ee

For Step 1, we can take advantage of the fact that the equations \eqref{3dN_odd_lambda}, \eqref{3dN_odd_rho} and \eqref{3dN_odd_rho_sum}
are exactly the same as \eqref{3dN_even_lambda}, \eqref{3dN_even_rho} and \eqref{3dN_even_rho_sum}
for $\scN$ even, with the only difference being that $i$ runs from $1$ to $(\scN-1)/2$.
We can thus simply reuse the result \eqref{3d_even_iso}; there is essentially no difference between $\scN$ even and $\scN$ odd,
 as far as Step 1 is concerned.

For next step, even though we have odd non-isotropic roots we can still follow the  improved algorithm (called Step 2$^{\prime}$ before),
since odd non-isotropic roots do not depend on the values of $\mathcal{N}$.
The analysis in this step is therefore the same as the $\scN=1$ case, with the only difference being the shift of $\Delta$
(compare the coefficients of $\beta_D$ and $\beta_J$ in \eqref{3dN_odd_rho_sum} and \eqref{3dN0_rho_sum}).

By combining these results, the module is reducible either at values \eqref{3d_even_iso} or
\begin{align}
\Delta=\left(\left\{ \frac{\ell}{2} +2-\frac{\scN}{2}-\mathbb{Z}_{>0}\right\} \setminus \left\{1-\frac{\scN}{2} \right\}\right)\bigcup \left(\frac{3}{2}-\frac{\scN}{2}-\mathbb{Z}_{>0}\right) \;.
\end{align}
for $\ell$ even and 
\begin{align}
\Delta=\left(\left\{ \frac{\ell}{2} +2-\frac{\scN}{2}-\mathbb{Z}_{>0}\right\} \setminus \left\{1-\frac{\scN}{2} \right\}\right)\bigcup \left(1-\frac{\scN}{2}-\mathbb{Z}_{>0}\right) \;.
\end{align}
for $\ell$ odd.

\section{\texorpdfstring{Summary of Results for $D=3, 4, 5, 6$}{Summary of Results for D=3, 4, 5, 6}}\label{sec.D}

Having worked out the case of three spacetime dimensions in detail, it is straightforward to 
repeat the analysis in spacetime dimensions four, five and six. 
The details of the computations are summarized in Appendix \ref{app.SCA}.
Let us here summarize the results for all the SCFTs. We hope this section will be useful to those readers who are interested only in the final results.

As in the rest of this paper, this list in general contains the same value of $\Delta$ multiple times.
In those cases it is an indication that the corresponding poles could be of second order of higher.

\subsection{\texorpdfstring{$D=3$}{D=3}}

Let $\ell\in \mathbb{Z}_{\ge 0}$ be twice the angular spin, and let us 
define $C_{\ell}$ to be $C_{\ell}=3$ for $\ell$ even, and $C_{\ell}=2$ for $\ell$ odd.
For $\scN=0$ (see \eqref{3dN0_result_1} and \eqref{3dN0_result_2})
\begin{align}
\left(\bigg\{ \frac{\ell}{2}+2-\mathbb{Z}_{>0}\bigg\} \setminus \bigg\{1-\frac{\ell}{2} \bigg\}\right)\bigcup \left(\frac{C_{\ell}}{2}-\mathbb{Z}_{>0}\right)\;.
\end{align}
For $\scN=1$ (see \eqref{3dN1_result_1}) 
\begin{align}
\left(\bigg\{ \frac{\ell}{2}+2-\mathbb{Z}_{>0}\bigg\} \setminus \bigg\{1,  1-\frac{\ell}{2}  \bigg\}\right)\bigcup \left(\frac{C_{\ell}}{2}-\mathbb{Z}_{>0}\right)\;.
\end{align} 

For $\scN$ even, 
\begin{align}
\begin{split}
&\bigcup_{a=1}^{\frac{\scN}{2}}\bigg\{
2-a+\ell+\lambda_a \;,
1-a-\ell+\lambda_a \;,
2-\scN+a+\ell-\lambda_a \;,
1-\scN+a-\ell-\lambda_a 
\bigg\} \\
&\qquad \bigcup\left(\left\{  \frac{\ell}{2}+2-\frac{\scN}{2}-\mathbb{Z}_{>0}\right\} \setminus \left\{1- \frac{\ell}{2}-\frac{\scN}{2} \right\}\right)\bigcup \left(\frac{C_{\ell}-\scN}{2}-\mathbb{Z}_{>0}\right) \;.
\end{split}
\end{align}
For $\scN$ odd
\begin{align}
\begin{split}
&\bigcup_{a=1}^{\frac{\scN-1}{2}}\bigg\{
2-a+\ell+k_a \;,
1-a-\ell+k_a \;,
2-\scN+a+\ell-\lambda_a \;,
1-\scN+a-\ell-\lambda_a 
\bigg\} \\
&\qquad \bigcup
\left(\left\{  \frac{\ell}{2}+2-\frac{\scN}{2}-\mathbb{Z}_{>0}\right\} \setminus \left\{1-\frac{\scN}{2}, 1- \frac{\ell}{2}-\frac{\scN}{2}\right\}\right)\bigcup \left(\frac{C_{\ell}-\scN}{2}-\mathbb{Z}_{>0}\right)
\end{split}
\end{align}

\subsection{\texorpdfstring{$D=4$}{D=4}}

Let us denote the angular spins of $\mathfrak{so}(4)$ by the Dynkin label $[\ell_1, \ell_2]$. 

For $\scN=0$ (see \eqref{4d})
\begin{align}
\Delta=\bigg\{ \frac{\ell_1+\ell_2}{2}+3-\mathbb{Z}_{>0}\bigg\} \setminus \bigg\{ \frac{|\ell_1-\ell_2|}{2}+2 \bigg\} \;.
\end{align}
For $\scN=1,2,3,4$ with $\mathfrak{su}(\scN)$ representation labeled by a partition $\{\lambda_a\}$: 
 \begin{align}
 \begin{split}
\Delta&=\left(\bigg\{ \frac{\ell_1+\ell_2}{2}+3-\scN-\mathbb{Z}_{>0}\bigg\} \setminus \bigg\{ \frac{|\ell_1-\ell_2|}{2}+2-\scN \bigg\} \right)  \\
&\bigcup_{a=1}^{\scN-1} \bigg\{ \ell_1+\frac{\scN-4}{2\scN} R   +2 \lambda_a-2\frac{|\lambda|}{\scN}\ -2a  +4\bigg\}  \\
&\bigcup_{a=1}^{\scN-1}   \bigg\{ -\ell_1+\frac{\scN-4}{2\scN}R   +2 \lambda_a-2\frac{|\lambda|}{\scN}\ -2a+2\bigg\}  \\
&\bigcup_{a=1}^{\scN-1}   \bigg\{  -\ell_2-\frac{\scN-4}{2\scN} R   -2 \lambda_a+2\frac{|\lambda|}{\scN}\ +2a+2-2\scN\bigg\} \\
&\bigcup_{a=1}^{\scN-1}  \bigg\{  \ell_2-\frac{\scN-4}{2\scN} R   -2 \lambda_a+2\frac{|\lambda|}{\scN}\ +2a-2\scN \bigg\}  \;,
\end{split}
 \end{align}
 where $|\lambda|$ is the size of the partition $\lambda$ \eqref{sumY}.
 For $\scN=4$ the dependence on $R$ drops out, as expected from the reduction of the symmetry algebra
 from $\mathfrak{sl}(2,2|4)$ to $\mathfrak{psl}(2,2|4)$.

\subsection{\texorpdfstring{$D=5$}{D=5}}

Let us denote the angular spins of $\mathfrak{so}(5)$ by the Dynkin label $[\ell_1, \ell_2]$. 

For $\scN=0$ (see \eqref{5d})
\begin{align}
\left(\bigg\{ \ell_1+\frac{\ell_2}{2}+4-\mathbb{Z}_{>0}\bigg\} \setminus \bigg\{\frac{\ell_2}{2}+3, -\frac{\ell_2}{2}+2 , -\ell_1-\frac{\ell_2}{2}+1\bigg\}\right)
\bigcup \left(\frac{5}{2}-\mathbb{Z}_{>0}\right) \;.
\end{align}
For $\scN=1$ with $\mathfrak{sp}(2)$ R-symmetry spin $k\in \mathbb{Z}_{\ge 0}$
\begin{align}
\begin{split}
&\bigg\{ \ell_1+\frac{\ell_2}{2}+3k+4,  \ell_1+\ell_2-3k+1,  \ell_1-\ell_2+3k+3, \ell_1-\ell_2-3k\bigg\} \\
&\qquad \bigcup \bigg\{-\ell_1+\ell_2+3k+1,  -\ell_1+\ell_2-3k-2, -\ell_1-\ell_2+3k, -\ell_1-\ell_2-3k-3\bigg\} \\
&\qquad\bigcup \left(\bigg\{ \ell_1+\frac{\ell_2}{2}+2-\mathbb{Z}_{>0}\bigg\} \setminus \bigg\{\frac{\ell_2}{2}+1, - \frac{\ell_2}{2}, -\ell_1- \frac{\ell_2}{2}-1\bigg\}\right)\\
& \qquad \bigcup \left(\frac{1}{2}-\mathbb{Z}_{>0}\right) \;.
\end{split}
\end{align}

\subsection{\texorpdfstring{$D=6$}{D=6}}

Let us denote the angular spins of $\mathfrak{so}(6)$ by the Dynkin label $[\ell_1, \ell_2, \ell_3]$. 

For $\scN=0$ (see \eqref{6d})
\begin{align}
\bigg\{ \ell_1+ \frac{\ell_2+\ell_3}{2}+5-\mathbb{Z}_{>0}\bigg\} \setminus\bigg\{\frac{\ell_2+\ell_3}{2}+4, \frac{|\ell_2-\ell_3|}{2}+3 \bigg\} \;.
\end{align}
For $\scN=(0,1)$ with $\mathfrak{sp}(2)\simeq \mathfrak{so}(3)$ R-symmetry spin $k\in \mathbb{Z}_{\ge 0}$ we have
\begin{align}
\begin{split}
&\bigg\{ 
\frac{ 2\ell_1 + \ell_2+3\ell_3}{2} + 2k +6,
\frac{ 2\ell_1 + \ell_2+3\ell_3}{2} + 2k +4
\bigg\} 
 \\
&\qquad\bigcup \bigg\{
\frac{ 2\ell_1+\ell_2 - \ell_3}{2} - 2k +4,
\frac{ 2\ell_1+\ell_2 - \ell_3}{2} + 2k+2
\bigg\}   \\
&\qquad\bigcup \bigg\{
\frac{ -2\ell_1+\ell_2 - \ell_3}{2} - 2k +2,
\frac{ -2\ell_1+\ell_2 - \ell_3}{2} + 2k
\bigg\}  \\
&\qquad\bigcup\bigg\{ 
\frac{ -2\ell_1-3\ell_2 - \ell_3}{2} + 2k ,
\frac{ -2\ell_1-3\ell_2 - \ell_3}{2} + 2k -2
\bigg\}
 \\
&\qquad\bigcup \left(\bigg\{ \ell_1+ \frac{\ell_2+\ell_3}{2}+3-\mathbb{Z}_{>0}\bigg\} \setminus\bigg\{\frac{\ell_2+\ell_3}{2}+2, \frac{|\ell_2-\ell_3|}{2}+1 \bigg\}\right)\;.
\end{split}
\end{align}
For $\scN=(0,2)$ with $\mathfrak{sp}(4) \simeq \mathfrak{so}(5)$ R-symmetry spin  labeled by Dynkin label $[k_1, k_2]$ we have
{\small
\begin{align}
\begin{split}
&\bigg\{ 
\frac{ 2\ell_1 + \ell_2+3\ell_3}{2} + 2k_1 +2k_2+6,
\frac{ 2\ell_1 + \ell_2+3\ell_3}{2} + 2k_1 -2k_2+4
\bigg\} 
 \\
&\qquad\bigcup \bigg\{
\frac{ 2\ell_1 + \ell_2+3\ell_3}{2} - 2k_1 +2k_2+0,
\frac{ 2\ell_1 + \ell_2+3\ell_3}{2} + 2k_1 -2k_2-2
\bigg\} 
 \\
&\qquad\bigcup\bigg\{ 
\frac{ 2\ell_1+\ell_2 - \ell_3}{2} + 2k_1 +2k_2+4,
\frac{ 2\ell_1+\ell_2 - \ell_3}{2} + 2k_1 -2k_2+2
\bigg\}
 \\
&\qquad\bigcup \bigg\{
\frac{ 2\ell_1+\ell_2 - \ell_3}{2} - 2k_1 +2k_2-2,
\frac{ 2\ell_1+\ell_2 - \ell_3}{2} + 2k_1 -2k_2-4
\bigg\}  \\
&\qquad\bigcup\bigg\{ 
\frac{ -2\ell_1+\ell_2 - \ell_3}{2} + 2k_1 +2k_2+2,
\frac{ -2\ell_1+\ell_2 - \ell_3}{2} + 2k_1 -2k_2
\bigg\}
 \\
&\qquad\bigcup \bigg\{
\frac{ -2\ell_1+\ell_2 - \ell_3}{2} - 2k_1 +2k_2-4,
\frac{ -2\ell_1+\ell_2 - \ell_3}{2} + 2k_1 -2k_2-6
\bigg\}  \\
&\qquad\bigcup\bigg\{ 
\frac{ -2\ell_1-3\ell_2 - \ell_3}{2} + 2k_1 +2k_2,
\frac{ -2\ell_1-3\ell_2 - \ell_3}{2} + 2k_1 -2k_2-2
\bigg\}
 \\
&\qquad\bigcup \bigg\{
\frac{-2\ell_1-3\ell_2 - \ell_3}{2} - 2k_1 +2k_2-6,
\frac{ -2\ell_1-3\ell_2 - \ell_3}{2} + 2k_1 -2k_2-8
\bigg\}  
\\
&\qquad \bigcup\left(\bigg\{ \ell_1+ \frac{\ell_2+\ell_3}{2}+1-\mathbb{Z}_{>0}\bigg\} \setminus\bigg\{\frac{\ell_2+\ell_3}{2}, \frac{|\ell_2-\ell_3|}{2}-1 \bigg\} \right)
 \;.
\end{split}
\end{align}
}

\section{Comments on Unitarity Bounds}\label{sec.unitarity}

Our analysis of the irreducibility of parabolic Verma modules, or equivalently 
poles of superconformal blocks, is closely related with the discussion of unitarity bounds for superconformal field theories~\cite{Flato:1983te,Dobrev:1985qv,Dobrev:1985vh,Dobrev:1985qz,Dobrev:2002dt,Minwalla:1997ka,Bhattacharya:2008zy,Buican:2016hpb,Cordova:2016emh}.

The theory is unitarity when the value of $\Delta$ is sufficiently large. As we lower the value of $\Delta$,
at some value $\Delta=\Delta_*$, we will encounter the null state, and the theory is no longer unitary 
at $\Delta=\Delta_*-\epsilon$ with $\epsilon>0$, at least when $\epsilon$ is chosen sufficiently small.
In other words, $\Delta_*$ is the minimal value so that the theory is always unitary (irreducible) as long as $\Delta>  \Delta_*$.
Such a value is sometimes known as the ``first reduction point'' in the literature.

It is straightforward to derive this first reduction point from our results summarized in Section \ref{sec.D}---$\Delta_*$ is simply the largest value of $\Delta$ which appears in the list of poles in Section \ref{sec.D}. The result is summarized in Table \ref{reduction_point}.

\begin{table}[htbp]
\centering
\caption{The values of the first reduction point $\Delta=\D_*$ for $\mathcal{N}$-extended superconformal algebra in $D=3,4,5,6$. We here assumed that the R-charges are generic,
e.g. $\lambda_1>0$ for $D=3, \scN>1$.}
\resizebox{.95\columnwidth}{!}{%
{\setlength{\extrarowheight}{6pt}
\begin{tabular}{|c||c|c|c|}
\hline
dim. & susy & $\{\ell_i\}=0$ & $\{\ell_i\}\neq0$\\
\hline
\hline
& $\mathcal{N}=0$ & $\D=\frac{1}{2}$ & $\D=\frac{\ell}{2}+1$\\
$D=3$ & $\mathcal{N}=1$ & $\D=\frac{1}{2}$ & $\D=\frac{\ell}{2}+1$\\

& $\mathcal{N}>1$ & $\D=1+\l_1$ & $\D=1+\frac{\ell}{2}+\l_1$\\
\hline
& $\mathcal{N}=0$ & $\D=1$ & $\D=2+\frac{\ell_1+\ell_2}{2}$\\
 & $\mathcal{N}=1$ & $\D=2+\frac{3R}{2}$ & $\D=\begin{cases}\ell_2+\frac{3R}{2}\,,\ \ell_1=0\\ \ell_1+2-\frac{3R}{2}\,,\ \ell_2=0\end{cases}  $\\

$D=4$ & $\mathcal{N}=2$ & $\D=2+\l_1-\frac{R}{2}$ & $\D=\begin{cases}\ell_1+2+\l_1-\frac{R}{2}\,,\ \ell_2=0\\ \ell_2+\frac{R}{2}-\l_1\,,\ \ell_1=0\end{cases} $\\

& $\mathcal{N}=3$ & $\D=2+\frac{2}{3}(2\l_1-\l_2)-\frac{R}{6}$ & $\D=\begin{cases}\ell_1+2+\frac{2}{3}(2\l_1-\l_2)-\frac{R}{6}\,,\ \ell_2=0\\ \ell_2+\frac{R}{6}+\frac{2}{3}(\l_1-2\l_2)-2\,,\ \ell_1=0\end{cases} $\\

& $\mathcal{N}=4$ & $\D=2+\frac{1}{2}(3\l_1-\l_2-\l_3)$ & $\D=\begin{cases}\ell_1+2+\frac{1}{2}(3\l_1-\l_2-\l_3)\,,\ \ell_2=0\\ \ell_2-\frac{1}{2}(3\l_3-\l_1-\l_2)-2\,,\ \ell_1=0\end{cases} $\\
\hline
$D=5$ & $\mathcal{N}=0$ & $\D=\frac{3}{2}$ & $\D=\begin{cases}2+\frac{\ell_2}{2}\,, \ \ell_1=0\,,\ell_2\neq0\\3+\ell_1\,, \ \ell_2=0\,,\ell_1\neq0\\3+\ell_1+\frac{\ell_2}{2}\,, \ \ell_1,\ell_2\neq0 
\end{cases}$\\
 & $\mathcal{N}=1$ & $\D=3k+4$ & $\D=\ell_1+\frac{\ell_2}{2}+3k+4$\\
\hline
$D=6$& $\mathcal{N}=0$ & $\D=2$ & $\D=\begin{cases}3+\frac{\ell_2+\ell_3}{2}\,, \ \ell_1=0\,,\ell_2,\ell_3\neq0\\ 4+\ell_1+\frac{\ell_3}{2}\,, \ \ell_2=0\,,\ell_1,\ell_3\neq0\\
4+\ell_1+\frac{\ell_2}{2}\,, \ \ell_3=0\,, \ell_1,\ell_2\neq0\\
4+\ell_1\,,\ \ell_2,\ell_3=0\,,\ell_1\neq0\\
3+\frac{\ell_2}{2}\,, \ \ell_1,\ell_3=0\,,\ell_2\neq0\\
3+\frac{\ell_3}{2}\,, \ \ell_1,\ell_2=0\,, \ell_3\neq0\\
4+\ell_1+\frac{\ell_2+\ell_3}{2}\,,\ \ell_1,\ell_2,\ell_3\neq0\\
\end{cases}
$\\
 & $\mathcal{N}=1$ & $\D=2k+6$ & $\D=\ell_1+\frac{\ell_2+\ell_3}{2}+2k+6$\\
& $\mathcal{N}=2$ & $\D=2(k_1+k_2)+6$ & $\D=\ell_1+\ell_2+\ell_3+2(k_1+k_2)+6$\\
\hline
\end{tabular}
}
}
\label{reduction_point}
\end{table}

In some literature on superconformal field theories, the value of the first reduction  point $\Delta_*$ has often been derived by the analysis of 
the Kac's irreducibility criterion \cite{Kac_LNM}. However, 
as we have already commented in Section \ref{sec.irreducibility}, the Kac's criterion is different from the irreducibility criterion of \cite{Oshima:2016gqy} which in our opinion is more directly of relevant to the study of superconformal field theory, in that in the latter
we need to take into account the conditions as originating from non-isotropic roots. 

Having said that, it turns out that the actual values of $\Delta_*$ obtained here is after all consistent with the values known in the literature, see e.g. \cite{Cordova:2016emh} for convenient reference. 
More detailed analysis will be needed to derive the full unitarity bounds along these lines, i.e.\ to classify unitary short multiplets.
The details will be left for future work, but as a small consistency check we can verify that all the known unitary short multiplets of the superconformal algebras have values of the scaling dimension $\Delta$ where the generalized Verma module is reducible.



We can further study more general short multiplets of the superconformal algebra---almost all such short multiplets are non-unitary.
In this case there is a propri no guarantee that the Kac's criterion and 
our irreducibility criterion give the same set of non-unitary short multiplets.
Since such multiplets could be of practical interest (e.g.\ in applications to statistical mechanics, just like
minimal models in the case of two-dimensional conformal field theories), 
it would be of fundamental importance to further study the 
(in general non-unitary) representations of superconformal algebras, perhaps along the lines of this paper.



\section{Further Implications for Superconformal Blocks}\label{sec.remarks}

The pole structures derived in this paper is completely general,
and can be applied to correlators of any operators in general representations.
Namely the only possible poles of the superconformal blocks as a function of the 
scaling dimension $\Delta$ should be exhausted by the results presented above.

Having said that, if we consider a superconformal block for a particular operators
then some of these poles might be absent. Such a phenomenon is 
already known for non-supersymmetric CFTs in three spacetime dimensions,
where the conformal block for four external scalar operators
does not have poles corresponding to the type IV null states in \cite{Penedones:2015aga}.
Such an absence of poles happens since the residues at these poles can vanish.
This in turn can happen because of the absence of the relevant three-point functions,
see \cite{Costa:2011mg,Kravchuk:2016qvl,Schomerus:2016epl}
for systematic analysis of possible three-point structures.

We would like to urge the readers to 
compare the predictions in this paper with 
their favorite expressions for superconformal blocks.
We leave the detailed such analysis to be future work,
but let us here make one small comment.

We have seen that in the ``improved algorithm'' of Section \ref{subsec.improved}
involving shifting the scaling dimension $\Delta$ by the number of supersymmetries $\scN$,
as far as the pole structure of superconformal blocks is concerned.
This is known to hold for superconformal blocks with external scalar operators.
Indeed, in four dimensions ($D=4$) the scalar superconformal block 
in theories of $\scN$-extended supersymmetry ($\scN=1,2,4$) is known to be of the form\cite{Fitzpatrick:2014oza} (see also \cite{Dolan:2001tt,Poland:2010wg})
\beq
\label{g_SUSY}
g^{\mathcal{N}}(u,v)=u^{-\frac{\mathcal{N}}{2}} g^{\Delta_{12}=\Delta_{34}=\mathcal{N}}_{\Delta+\mathcal{N}, l}(u, v) \ . 
\eeq
where $g_{\Delta,l}$ is the corresponding quantity for the case of $\scN=0$ theory\cite{Dolan:2000ut,Dolan:2003hv}
\beq
g_{\Delta,l}(z, \bar{z})=
(-1)^l
\frac{z\bar{z}}{z-\bar{z}}
\left[
k_{\Delta+l}(z) k_{\Delta-l-2}(\bar{z})
-
k_{\Delta+l}(\bar{z}) k_{\Delta-l-2}(z) 
\right] \;,
\eeq
where the different parametrizations of the cross-ratio coordinates are related by
\beq
u=z\bar{z} \  ,\quad
v=(1-z)(1-\bar{z}) \;,
\eeq
and 
\beq
k_{\beta}(x):= 
x^{\frac{\beta}{2}} \, _2 F_1\left(\frac{\beta}{2},\frac{\beta}{2}, \beta; x\right)
=x^{\frac{\beta}{2}} 
\sum_{n=0}^{\infty}
\frac{\left[(\frac{\beta}{2})_n\right]^2}{(\beta)_n n!} x^n 
 \ ,
\eeq
with the hypergeometric function $_2 F_1(a,b,c;x)$ and the Pochhammer symbol $(x)_n$ defined by
\begin{align}
& _2 F_1(a,b,c;x)=
\sum_{n=0}^{\infty}
\frac{(a)_n (b)_n}{(c)_n n!} x^n  \ , \\
& (x)_0:=1 \ , \quad (x)_n:=x (x+1) (x+2) \ldots (x+n-1) \ .
\end{align}

In this example that effect of $\scN$ is precisely to shift the positions of poles in $\Delta$,
as one expected from the analysis of Section \ref{subsec.improved}.\footnote{Of course, for complete analysis one also needs to take into account
possible poles \eqref{4dN_iso}, as originating from isotropic roots, which are not shifted simply by $\scN$.}
Interestingly, a similar pattern is known to hold more generally in theories of four supercharges in other (in general non-integer) dimensions, 
see \cite{Bobev:2015jxa}.\footnote{It is known that such a relation does not hold for theories with eight supercharges, see \cite{Bobev:2017jhk}.
We thank Nikolay Bobev for discussions related to this point.} This could be an indication that the pole structures as derived in this paper might be 
more constraining as one might have expected for a generic function.

One can use the pole structure studied in this paper 
as a input to the recursion relations of superconformal blocks,
along the lines of \cite{Penedones:2015aga}.
The completion of this program requires explicit expressions for the null states and their norms (when $\Delta$ is generic),
which seems to be unknown in the literature. It would be interesting to explore this further.
We will leave these questions for future work.

\section*{Acknowledgments}

We would like to thank Yoshiki Oshima for related discussion. 
The bulk of this project was completed in fall 2016--spring 2017, however the publication has been delayed for no good reason.
We thank the organizers of the symposium ``Bootstrap Approach to Conformal Field Theories and Applications'', OIST, March 2018,
for providing motivation to finish up this work.
The authors are supported in part by WPI program (MEXT, Japan). MY is also supported by by JSPS Program for Advancing Strategic International Networks to Accelerate the Circulation of Talented Researchers, by JSPS Grant No.\ 15K17634 and No.\ 17KK0087, and by JSPS-NRF Joint Research Project. He would like to thank Harvard university for hospitality.

\appendix

\section{\texorpdfstring{$ABD$ Roots and Weights}{ABD Roots and Weights}}\label{app.ABD}

In this Appendix we summarize some standard facts about the root system and the highest weights for the 
$ABD$ Lie algebras. We can spare the $C$ case (associated with Lie algebra $\mathfrak{sp}(2N)$) for the applications to SCFTs in this paper, since we can
use the isomorphism of Lie algebras: $\mathfrak{sp}(2)=\mathfrak{su}(2)$, $\mathfrak{sp}(4)=\mathfrak{so}(5)$.

\subsection{\texorpdfstring{$A_{N-1}$}{A(N-1)}}\label{app.A_root}

The $A_{N-1}$ root system is given by
\begin{align}
&\Delta_{\bar{0}}=\bigg\{ 
\pm (\delta_a - \delta_b)
\bigg\} \ ,
\end{align}
where we  introduced a orthonormal basis $\delta_a$
\beq
\label{ortho}
(\delta_a, \delta_b)=\delta_{a, b} \ ,  \quad a, b=1, \ldots, N \;.
\eeq
Under an ordering $\delta_1 > \dots >\delta_{\scN}$,
the positive simple roots are given by 
$\alpha_a=\delta_a-\delta_{a+1}$, with $a=1, \ldots, N-1$.

The fundamental weight is given by
\begin{align}
w_a:=\sum_{b=1}^{a} \delta_b -\frac{a}{N} \sum_{c=1}^{N} \delta_c \;, \quad
\alpha_a:=\delta_a-\delta_{a+1} \;,
\end{align}
which are determined by the conditions
\begin{align}
&2\frac{ (w_a, \alpha_b)}{(\alpha_b, \alpha_b)}=\delta_{ab} \;, \quad a,b=1, \dots, N \;, \quad
\alpha_a:=\delta_a-\delta_{a+1} \;,
\label{fundamental}
\end{align}
as well as the condition that $w_a$ is in the $\mathbb{R}$-span of $\alpha_b$'s.

The highest weight for a finite-dimensional representation of the $\mathfrak{su}(N)$ algebra is given by 
a dominant integral weight, which is given by a $\mathbb{Z}_{\ge 0}$-span of fundamental weights:
\be
\lambda=\sum_{a=1}^{N-1} \ell_a w_a
=\sum_{a=1}^{N-1} \ell_a \left(\sum_{b=1}^{a} \delta_b -\frac{a}{N} \sum_{c=1}^{N} \delta_c \right) \;,
\label{A_lambda}
\ee
where the coefficients $[\ell_1, \dots, \ell_{\scN}]$ are called Dynkin labels.
The highest weight \eqref{A_lambda} can also be written as
\be
\lambda=\sum_{a=1}^{N-1} \ell_a w_a
=\sum_{a=1}^{N} \left(\lambda_a  -\frac{|\lambda|}{N}  \right) \delta_a  \;,
\ee
where we introduced $\lambda_a$ and $|\lambda|$ by
\begin{align}
&\lambda_a=\ell_a+ \ell_{a+1}  +\dots +\lambda_{N-1} \;, \quad
\textrm{or} \quad \ell_a=\lambda_a-\lambda_{a+1} \;,\\
&|\lambda|=\sum_a \lambda_a \label{sumY} \;.
\end{align}
Since this $\lambda$ satisfies
\begin{align}
\lambda_1 \ge \lambda_2 \ge \ldots \ge \lambda_{N-1} \ge 0 \;,
\end{align}
$\{ \lambda_a \}$ define a partition (Young diagram), where the number of boxes at height $a$ 
is given by $\lambda_a$; in this language, $|\lambda|$ is the total number of boxes.

\subsection{\texorpdfstring{$D_{N}$}{D(N)}}\label{app.D_root}

The $D_N$ root system (corresponding to the Lie algebra $\mathfrak{so}(2N)$) is given by
\begin{align}
&\Delta_{\bar{0}}=\bigg\{ 
\pm \delta_a \pm \delta_b
\bigg\} \ ,  \quad a,b=1, \dots, N \;,
\end{align}
and the inner product is given by \eqref{ortho}.
In dictionary ordering $\delta_1> \delta_2>\ldots>\delta_N$, the positive simple roots are given by
\begin{align}
\alpha_a=\delta_a-\delta_{a+1} \quad a=1, \dots, N-1 \;, \quad
\alpha_N=\delta_{N-1}+\delta_N \;.
\end{align}
The fundamental weights satisfying \eqref{fundamental} are obtained as
\begin{align}
\begin{split}
w_a &=\delta_1 + \dots + \delta_a \;, \quad a=1\;, \dots, N-2 \;, \\
w_{N-1}&=\frac{\delta_1 + \dots + \delta_{N-1}-\delta_N}{2} \;,\\
w_{N}&=\frac{\delta_1 + \dots + \delta_{N-1}+\delta_N}{2} \;.\\
\end{split}
\end{align}
The highest weight for a finite-dimensional representation is given as
\beq
\lambda=\sum_{a=1}^N \ell_a  w_a
=\sum_{a=1}^N \lambda_a  \delta_a
\eeq
where $\ell_a\in \mathbb{Z}_{\ge 0}$ are integers called Dynkin labels, and we defined half-integers $\lambda_a\in \mathbb{Z}_{\ge 0}/2$ by 
\begin{align}\label{lambda_delta_even}
\begin{split}
&\lambda_a = \ell_a+ \dots \ell_{N-2}+\frac{\ell_{N-1}+\ell_N}{2} \;, \quad a=1, \dots, N-2 \;, \\
&\lambda_{N-1}=\frac{\ell_{N-1}+\ell_N}{2} \;, \quad 
\lambda_{N}=\frac{-\ell_{N-1}+\ell_N}{2} \;.
\end{split}
\end{align}
By definition this satisfies $\lambda_1 \ge \lambda_2 \ge \dots \ge \lambda_{N-1}\ge |\lambda_N| \ge 0$.
Note that $\lambda_N$ can be negative.
The $\lambda_a$'s are half-integers for a general representation. If we consider the tensor representation, however,
all the $\lambda_a$'s takes values in integers; a tensor representation is labeled by a partition and a sign.

\subsection{\texorpdfstring{$B_{N}$}{B(N)}}\label{app.B_root}

The $B_N$ root system (corresponding to the Lie algebra $\mathfrak{so}(2N+1)$) is similar to the $D_N$ root system, but 
with some extra roots added:
\begin{align}
&\Delta_{\bar{0}}=\bigg\{ 
\pm \delta_a \pm \delta_b \;,
\pm \delta_a
\bigg\}  \;, \quad a,b=1, \dots, N
\end{align}
and the inner product is given by \eqref{ortho}.
In dictionary ordering $\delta_1> \delta_2>\ldots>\delta_N$, the positive simple roots are given by
\begin{align}
\alpha_a=\delta_a-\delta_{a+1} \quad a=1, \dots, N-1 \;, \quad
\alpha_N=\delta_N \;.
\end{align}
The fundamental weights satisfying \eqref{fundamental} are obtained as
\begin{align}
\begin{split}
w_a &=\delta_1 + \dots + \delta_a \;, \quad a=1\;, \dots, N-1 \;, \\
w_{N}&=\frac{\delta_1 + \dots + \delta_{N-1}+\delta_N}{2} \;.\\
\end{split}
\end{align}
The highest weight for a finite-dimensional representation is given as
\beq
\lambda=\sum_{a=1}^N \ell_a  w_a
=\sum_{a=1}^N \lambda_a  \delta_a
\eeq
where $\ell_a\in \mathbb{Z}_{\ge 0}$ are Dynkin labels and we defined half-integers $\lambda_a\in \mathbb{Z}_{\ge 0}/2$ by 
\begin{align}\label{lambda_delta_odd}
\begin{split}
&\lambda_a = \ell_a+ \dots \ell_{N-2}+\frac{\ell_N}{2} \;, \quad a=1, \dots, N-1 \;, \\
&\lambda_{N}=\frac{\ell_N}{2}  \;.
\end{split}
\end{align}
By definition this satisfies $\lambda_1 \ge \lambda_2 \ge \dots \ge \lambda_N\ge 0$.
In tensor representations all the $\lambda_a$'s are integers; a tensor representation is 
labeled by a partition.

\section{\texorpdfstring{$\mathcal{N}=0$ Results}{N=0 Results}}\label{app.N0}

In this appendix we present the analysis for the non-supersymmetric cases (i.e.\ $\mathcal{N}=0$)
in spacetime dimensions $4, 5$ and $6$. Note that in these cases
the conformal algebra contains no odd elements,
and hence $\Delta =\Delta_{\bar{0}}, \Delta_1=\Delta_{1}^{+}=\varnothing$ and $\rho=\rho_0$.

Note that such an analysis (for a general spacetime dimension) was already given in \cite{Penedones:2015aga}, 
except there the spinor representations are not considered there. In this appendix we therefore present 
the full analysis including the cases of spinor representations.
We present this analysis in the same notations as in the rest of this paper, to make the comparison easier.\footnote{Please be aware of the notation change: $\ell_a$ in \cite{Penedones:2015aga} is denoted as 
$\lambda_a$ in this paper.}
We will find that while intermediate steps of the analysis changes slightly from \cite{Penedones:2015aga}, the 
final result is in the end the same as in  \cite{Penedones:2015aga}, even for spinor representations.

\subsection{\texorpdfstring{4d $\scN =0$}{4d N=0}} \label{app.D4N0}

The conformal algebra in this case is  $\mathfrak{g}=\mathfrak{su}(4)=A(3)$.
Let us introduce a basis
\beq
(\varepsilon_i, \varepsilon_j)=\delta_{i,j} \ , \quad
\eeq
with $i,j=1, \ldots, 4$.
Then we have
\begin{align}
&\Delta=\bigg\{ 
\pm( \varepsilon_i - \varepsilon_j )
\bigg\} \ , \\
&\Delta_{\fl}=\bigg\{ 
\pm( \varepsilon_1 - \varepsilon_2 ),   \quad
\pm( \varepsilon_3 - \varepsilon_4)
\bigg\}\;.
\end{align}
Let us choose an ordering $\varepsilon_1 > \varepsilon_2 >-\varepsilon_3>-\varepsilon_4$.
We then have
\begin{align}
\Delta^+&=\bigg\{
\varepsilon_{1}-\varepsilon_{2}  \;, \quad
\varepsilon_{4}-\varepsilon_{3}  
\bigg\}\;, \\
\Delta_{\fn}&=\bigg\{
\varepsilon_1 - \varepsilon_3 ,   \quad
\varepsilon_1 - \varepsilon_4 ,   \quad
\varepsilon_2 - \varepsilon_3 ,   \quad
\varepsilon_2 - \varepsilon_4 
 \bigg\}\;.
\end{align}

The highest weight vector and the Weyl vector is
\begin{align}
\begin{split}
\l&=-\frac{\D}{2}(\varepsilon_1+\varepsilon_2-\varepsilon_3-\varepsilon_4)+\frac{\ell_1}{2}(\varepsilon_1-\varepsilon_2)-\frac{\ell_2}{2}(\varepsilon_3-\varepsilon_4) \\
\r &=\frac{1}{2} \left(3 \varepsilon_1 +\varepsilon_2 -3\varepsilon_3 -\varepsilon_4 \right) \,,\\
\l+\rho&=\frac{-\Delta+3}{2} (\varepsilon_1 -\varepsilon_3 ) +\frac{-\Delta+1}{2}(\varepsilon_2 -\varepsilon_4)  
+\frac{\ell_1}{2}(\varepsilon_1-\varepsilon_2)-\frac{\ell_2}{2}(\varepsilon_3-\varepsilon_4)
\,.
\end{split}
\end{align}
Here $\ell_1$ and $\ell_2$ are the two angular spins, and take integer values.

We find
\begin{align}
\begin{split}
&n_{\varepsilon_1-\varepsilon_3}=3-\D+\frac{\ell_1+\ell_2}{2}\,, \ \ n_{\varepsilon_1-\varepsilon_4}=2-\D+\frac{\ell_1-\ell_2}{2}\,,\\
&n_{\varepsilon_2-\varepsilon_3}=2-\D+\frac{\ell_2-\ell_1}{2}\,, \ \ n_{\varepsilon_2-\varepsilon_4}=1-\D-\frac{\ell_1+\ell_2}{2}\,.
\end{split}
\end{align}
and $\Psi_\l^+\neq\varnothing$ in the following cases:
\be
\D-C_{\ell_1+\ell_2} =
\begin{dcases}
3+\frac{\ell_1+\ell_2}{2}-k\,, \ k=\{1,\dots\infty\}\,,\ \Psi_\l^+=\{\e_1-\e_3\}\\
2-\frac{\ell_1-\ell_2}{2}-k\,,\  k=\{1,2,\dots,1+\ell_2\}\,,\ \Psi_\l^+=\{\e_1-\e_3,\e_1-\e_4,\e_2-\e_3\}\,,\\
2+\frac{\ell_1-\ell_2}{2}-k\,, \ k=\{1,2,\dots\ell_1-\ell_2\}\,, \Psi_\l^+=\{\e_1-\e_3,\e_1-\e_4\}\,,\\
1-\frac{\ell_1+\ell_2}{2}-k\,, \ k=\{1,2,\dots\}\,, \Psi_\l^+=\{\e_1-\e_3,\e_1-\e_4,\e_2-\e_3,\e_2-\e_4\}\,,\\
\end{dcases}
\ee
for $\ell_1\ge \ell_2$, and 
\be
\D-C_{\ell_1+\ell_2} =\begin{dcases}1-\frac{\ell_1+\ell_2}{2}-k\,, \ k=\{1,2,\dots\}\,, \Psi_\l^+=\{\e_1-\e_3,\e_1-\e_4,\e_2-\e_3,\e_2-\e_4\}\,,\\
2+\frac{\ell_2-\ell_1}{2}-k\,, \ k=\{1,2,\dots,\ell_2-\ell_1\}\,, \Psi_\l^+=\{\e_1-\e_3,\e_2-\e_3\}\,,\\
2-\frac{\ell_2-\ell_1}{2}-k\,,\  k=\{1,2,\dots,1+\ell_1\}
\,,\ \Psi_\l^+=\{\e_1-\e_3,\e_1-\e_4,\e_2-\e_3\}\,,\\
3+\frac{\ell_1+\ell_2}{2}-k\,, \ k=\{1,\dots\infty\}\,,\ \Psi_\l^+=\{\e_1-\e_3\}\end{dcases}
\ee
for $\ell_1< \ell_2$.  Here we defined
\begin{align}
C_{\ell_1+\ell_2}=
\begin{dcases}
0   & (\ell_1+\ell_2\textrm{ even})\\
\frac{1}{2} & (\ell_1+\ell_2 \textrm{ odd})
\end{dcases}
\end{align}

The final step is to check whether there exists $\b\in\D_\mathfrak{n}$ such that $(\l+\r,\b)=0$. This happens for
\begin{align}
\begin{split}
&\D=3+\frac{\ell_1+\ell_2}{2}\,, \ n_{\varepsilon_{1}-\varepsilon_{3}}=0\,, \ \Psi_\l^+=\varnothing\,.\\
&\D=2+\frac{\ell_1-\ell_2}{2}\,, \ n_{\varepsilon_{1}-\varepsilon_{4}}=0\,, \ \Psi_\l^+=\begin{dcases}\{\e_1-\e_3,\e_2-\e_3\}\,,\ell_2>\ell_1\\ \{\e_1-\e_3\}\,,\ell_2<\ell_1\end{dcases}\,.\\
&\D=2+\frac{\ell_2-\ell_1}{2}\,, \ n_{\varepsilon_{2}-\varepsilon_{3}}=0\,,\ \Psi_\l^+=\begin{dcases}\{\e_1-\e_3,\e_1-\e_4\}\,,\ell_1>\ell_2\\ \{\e_1-\e_3\}\,,\ell_1<\ell_2\end{dcases}\,.\\
&\D=1-\frac{\ell_1+\ell_2}{2}\,,\ n_{\varepsilon_{2}-\varepsilon_{4}}=0\,, \ \Psi_\l^+=\{\e_1-\e_3,\e_1-\e_4,\e_2-\e_3\}\,.
\end{split}
\end{align}

The first line is trivial since $\Psi_\l^+=\varnothing$. For the remaining cases, we need to apply the final condition \eqref{eq.final}.
In most cases the condition \eqref{eq.final} does not hold.
The first exception is the obvious case of $\D=3+\frac{\ell_1+\ell_2}{2}$, when $\Psi_\l^+$ is empy.
The other exception, assuming $\ell_1>\ell_2$,
happens for $\D=2+\frac{\ell_1-\ell_2}{2}$.
In this case $\lambda+\rho$ is in the hyperplane orthogonal to $\varepsilon_1 -\varepsilon_4$,
and hence $s_{\e_1-\e_3}(\lambda+\rho)$ is orthogonal to $\varepsilon_3 -\varepsilon_4$.
This means that $s_{\e_1-\e_3}(\lambda+\rho)$  is fixed by an element of the Weyl group $W_{\mathfrak{l}}$
exchanging  $\varepsilon_3$ and $\varepsilon_4$.
The case of $\ell_1<\ell_2$ is similar, and we find
 the reducible points in four dimensions comprises of,
\be
\D=\bigg\{3+\frac{\ell_1+\ell_2}{2}-\mathbb{Z}_{>0}\bigg\}\bigg\backslash\bigg\{2+\frac{|\ell_1-\ell_2|}{2}\bigg\}\,
\label{4d}
\ee
In terms of partitions $\lambda_1, \lambda_2\in \mathbb{Z}$ with $\lambda_1\ge |\lambda_2|\ge 0$ (see Appendix \ref{app.ABD}),
this becomes
\begin{align}
\Delta=\bigg\{\frac{ \lambda_1}{2}+3-\mathbb{Z}_{>0}\bigg\} \setminus \bigg\{ \frac{|\lambda_2|}{2}+2 \bigg\} \;.
\end{align}

\subsection{\texorpdfstring{5d $\mathcal{N}=0$}{5d N=0}}

For this case, we use the root system for $\mathfrak{so}_7=B(3)$:
\begin{align}
&\Delta=\bigg\{ 
\pm \beta_D \pm \beta_{J_1},  \quad
\pm \beta_D \pm \beta_{J_2},  \quad
\pm \beta_{J_1} \pm \beta_{J_2},  \quad
\pm \beta_D ,  \quad
\pm \beta_{J_1},  \quad
\pm \beta_{J_2}
\bigg\} \ ,
\end{align}
with the inner product $(\beta_i, \beta_j)=\delta_{i,j}$
Under an ordering $\beta_D> \beta_{J_1}>\beta_{J_2}$,
\begin{align}
&\Delta_{\fl}=\bigg\{ 
\pm \beta_{J_1} \pm \beta_{J_2},  \quad
\pm \beta_{J_1},  \quad
\pm \beta_{J_2}
\bigg\}\;, \\
&\Delta_{\fn}=\bigg\{ 
\beta_D \pm \beta_{J_1},  \quad
\beta_D \pm \beta_{J_2},  \quad
\beta_D 
 \bigg\}\;.
\end{align}
We have 
\begin{align}
&\lambda=-\Delta \beta_D + \left( \ell_1 +\frac{\ell_2}{2}\right) \beta_{J_1}+\frac{\ell_2}{2} \beta_{J_2} + k\delta\;,\\
&\rho=\frac{5}{2}\beta_D+\frac{3}{2}\beta_{J_1}+\frac{1}{2}\beta_{J_2}+\frac{1}{2}\delta \;,\\
&\lambda+\rho=\left(-\Delta+\frac{5}{2}\right)\beta_D+\left(\ell_1+\frac{\ell_2}{2}+\frac{3}{2}\right)\beta_{J_1}+\left(\frac{\ell_2}{2}+\frac{1}{2}\right)\beta_{J_2}+ \left( k+\frac{1}{2} \right)\delta \;.
\end{align}

In Step 1 we have
 \begin{align}
\begin{split}
&n_{\b_D}=5-2\D\,,\\
&n_{\b_D+\b_{J_1}}=4-\D+\ell_1+\frac{\ell_2}{2}\,, \\
&n_{\b_D-\b_{J_1}}=1-\D-\ell_1-\frac{\ell_2}{2}\,,\\
&n_{\b_D+\b_{J_2}}=3-\D+\frac{\ell_2}{2}\,,\\
&n_{\b_D-\b_{J_2}}=2-\D-\frac{\ell_2}{2}\,.
\end{split}
\end{align}
This means that we have $\Psi_\l^+\neq\varnothing$ when
\be
\ell_2 \text{ even}\,, \ \ \D=\begin{dcases}
1-\ell_1-\frac{\ell_2}{2}-k\,, \ \Psi_\l^+=\{\b_D\,,\b_D\pm\b_{J_1}\,,\b_D\pm\b_{J_2}\}\\
2-\frac{\ell_2}{2}-k\,, \ k=\{1,\dots, 1+\ell_1\}\ \Psi_\l^+=\{\b_D\,,\b_D+\b_{J_1}\,,\b_D\pm\b_{J_2}\}\\
\frac{5}{2}-k\,, \ k=\{\frac{3}{2},\dots,\frac{\ell_2+1}{2}\}\ \Psi_\l^+=\{\b_D\,,\b_D+\b_{J_1}\,,\b_D+\b_{J_2}\}\\
3+\frac{\ell_2}{2}-k\,, \ k=\{1,\dots,\frac{1+\ell_2}{2}\}\ \Psi_\l^+=\{\b_D+\b_{J_1}\,,\b_D+\b_{J_2}\}\\
4+\ell_1+\frac{\ell_2}{2}-k\,, \ k=\{1,\dots,1+\ell_1\}\ \Psi_\l^+=\{\b_D+\b_{J_1}\} \\
\frac{5}{2}-k\,, \ k=\{1,\dots,\frac{\ell_2}{2}\}\ \Psi_\l^+=\{\b_D \}\\
\end{dcases}
\ee
and,
\be
\ell_2\text{ odd}\,, \ \ \D=\begin{dcases}
1-\ell_1-\frac{\ell_2}{2}-k\,, \ \Psi_\l^+=\{\b_D\,,\b_D\pm\b_{J_1}\,,\b_D\pm\b_{J_2}\}\\
2-\frac{\ell_2}{2}-k\,, \ k=\{\frac{1}{2},\dots, \ell_1+\frac{1}{2}\}\ \Psi_\l^+=\{\b_D\,,\b_D+\b_{J_1}\,,\b_D\pm\b_{J_2}\}\\
\frac{5}{2}-k\,, \ k=\{\frac{1}{2},\dots,\frac{\ell_2}{2}\}\ \Psi_\l^+=\{\b_D\,,\b_D+\b_{J_1}\,,\b_D+\b_{J_2}\}\\
3+\frac{\ell_2}{2}-k\,, \ k=\{\frac{1}{2},\dots,\frac{\ell_2}{2}\}\ \Psi_\l^+=\{\b_D+\b_{J_1}\,,\b_D+\b_{J_2}\}\\
4+\ell_1+\frac{\ell_2}{2}-k\,, \ k=\{\frac{1}{2},\dots,\ell_1+\frac{1}{2}\}\ \Psi_\l^+=\{\b_D+\b_{J_1}\}
\end{dcases}
\ee
The next step is to identify the walls for which there exists $\b\in\D_\mathfrak{n}$ such that $(\l+\r,\b)=0$,
\begin{align}
\begin{split}
&n_{\b_D}=0\,,  \ \D=\frac{5}{2}\,, \ \Psi_\l^+=\{\b_D+\b_{J_1}\,,\b_D+\b_{J_2}\}\\
&n_{\b_D+\b_{J_1}}=0\,, \ \D=4+\ell_1+\frac{\ell_2}{2}\,, \ \Psi_\l^+=\varnothing\\
&n_{\b_D-\b_{J_1}}=0\,, \ \D=1-\ell_1-\frac{\ell_2}{2}\,, \ \Psi_\l^+=\{\b_D\,,\b_D+\b_{J_1}\,,\b_D\pm\b_{J_2}\}\\
&n_{\b_D+\b_{J_2}}=0\,,\ \D=3+\frac{\ell_2}{2}\,, \ \Psi_\l^+=\{\b_D+\b_{J_1}\}\\
&n_{\b_D-\b_{J_2}}=0\,, \ \D=2-\frac{\ell_2}{2}\,, \ \Psi_\l^+=\{\b_D\,,\b_D+\b_{J_1}\,,\b_D+\b_{J_2}\}\,.
\end{split}
\end{align}
Finally we need to check the condition \eqref{eq.final}
for each of the above cases. In most cases the condition is not satisfied and the representation is 
reducible. The exception in the case $n_{\b_D}=0$, when $\Psi_\l^+$ has two elements $\b_D+\b_{J_1}\,,\b_D+\b_{J_2}$,
and $s_{\b_D+\b_{J_1}}(\lambda+\rho)$ and $s_{\b_D+\b_{J_2}}(\lambda+\rho)$ are each in the hyperplane orthogonal to 
$\b_{J_1}$ and $\b_{J_2}$. Since these two vectors can be rotated by an element of $W_{\mathfrak{l}}$ (rotation symmetry)
we have an irreducible representation.
Hence we get the following set of reducible points,
\be
\D=\bigg(\bigg\{4+\ell_1+\frac{\ell_2}{2}-\mathbb{Z}_{>0}\bigg\}\bigg\backslash\bigg\{3+\frac{\ell_2}{2}\,,2-\frac{\ell_2}{2}\,,1-\ell_1-\frac{\ell_2}{2}\bigg\}\bigg)\bigcup\bigg(\frac{5}{2}-\mathbb{Z}_{>0}\bigg)\,.
\label{5d}
\ee
In terms of $\lambda_1 \ge \lambda_2 \ge 0$ 
this can be written as
\begin{align}
\begin{split}
&\Delta=\left(\bigg\{ \lambda_1+4-\mathbb{Z}_{>0}\bigg\} \setminus \bigg\{\lambda_2+3, -\lambda_2+2 , -\lambda_1+1\bigg\}\right)
\bigcup \left(\frac{5}{2}-\mathbb{Z}_{>0}\right)  \;.
\end{split}
\end{align}

\subsection{\texorpdfstring{6d $\mathcal{N}=0$}{6d N=0}}

The root system for $\mathfrak{g}=\mathfrak{so}(8)=D(4)$ is
\begin{align}
&\Delta_{\bar{0}}=\bigg\{ 
\pm \alpha_i \pm \alpha_j,   \bigg\} \quad (i,j=D,1,2,3)\;, 
\end{align}
Under an ordering $\a_D>\a_1>\a_2>\a_3$,
\begin{align}
&\Delta_{\fl}=\bigg\{ 
\pm \alpha_1 \pm \alpha_2,  \quad
\pm \alpha_1 \pm \alpha_3,  \quad
\pm \alpha_2 \pm \alpha_3 \bigg\}\;,\\
&\Delta_{\fn}\cap \overline{\Delta}_0=\Delta_{\fn}=\bigg\{ \alpha_D\pm \alpha_{1}, \quad  \alpha_D\pm \alpha_{2}, \quad \alpha_D\pm \alpha_{3} \bigg\}\;,
\end{align}
with the constraint $s_1 s_2 s_3=1$.
{\small
\begin{align}
&\lambda= -\Delta \alpha_D +\left(\ell_1 +\frac{\ell_2+\ell_3}{2} \right) \alpha_1+\frac{\ell_2+\ell_3}{2}\alpha_2+\frac{-\ell_2+\ell_3}{2} \alpha_3+k \beta \;,\\
&\rho=3\alpha_D+2 \alpha_1+\alpha_2+\frac{1}{2}\beta\;, \\
&\lambda+\rho=( -\Delta+3) \alpha_D +\left(\ell_1 +\frac{\ell_2+\ell_3}{2}  +2\right)\alpha_1+\left(\frac{\ell_2+\ell_3}{2}  +1\right) \alpha_2+ \frac{-\ell_2+\ell_3}{2}   \alpha_3+\left(k+\frac{1}{2}\right) \beta\;.
\end{align}
}%
We compute
\begin{align}
\begin{split}
& n_{\a_D+\a_1}=5-\D+\ell_1+\frac{\ell_2+\ell_3}{2}\,,\\
& n_{\a_D-\a_1}=1-\D-\ell_1-\frac{\ell_2+\ell_3}{2}\,,\\
& n_{\a_D+\a_2}=4-\D+\frac{\ell_2+\ell_3}{2}\,,\\
& n_{\a_D-\a_2}=2-\D-\frac{\ell_2+\ell_3}{2}\,,\\
& n_{\a_D+\a_3}=3-\D-\frac{\ell_2-\ell_3}{2}\,,\\
& n_{\a_D-\a_3}=3-\D+\frac{\ell_2-\ell_3}{2}\,.
\end{split}
\end{align}
This means that if
\be
\D=5+\ell_1+\frac{\ell_2+\ell_3}{2}-\mathbb{Z}_{>0}\,,
\ee
then we will have $\Psi_\l^+\neq\varnothing$ with varying elements depending on what $n_\b$ are non-zero. We will directly go to the next step which is to analyze the wall condition for $\b\in\D_\mathfrak{n}$ such that $(\l+\r,\b)=0$:
\begin{align}
\begin{split}
& n_{\a_D+\a_1}=0\,,\ \D=5+\ell_1+\frac{\ell_2+\ell_3}{2}\,, \ \Psi_\l^+=\varnothing\,,\\
& n_{\a_D-\a_1}=0\,,\ \D=1-\ell_1-\frac{\ell_2+\ell_3}{2}\,, \ \Psi_\l^+=\{\a_D+\a_1\,,\a_D\pm\a_2\,,\a_D\pm\a_3\}\,,\\
& n_{\a_D+\a_2}=0\,,\ \D=4+\frac{\ell_2+\ell_3}{2}\,, \ \Psi_\l^+=\{\a_D+\a_1\}\,,\\
& n_{\a_D-\a_2}=0\,,\ \D=2-\frac{\ell_2+\ell_3}{2}\,, \ \Psi_\l^+=\{\a_D+\a_1\,,\a_D+\a_2\,,\a_D\pm\a_3\}\,,\\
& n_{\a_D+\a_3}=0\,,\ \D=3-\frac{\ell_2-\ell_3}{2}\,,\ \Psi_\l^+=\begin{cases}\{\a_D+\a_1\,,\a_D+\a_2\,,\a_D-\a_3\}\,, \ \ell_2>\ell_3\\
\{\a_D+\a_1\,,\a_D+\a_2\}\,, \ \ell_2<\ell_3\end{cases}\,,\\
& n_{\a_D-\a_3}=0\,,\ \D=3+\frac{\ell_2-\ell_3}{2}\,,\ \Psi_\l^+=\begin{cases}\{\a_D+\a_1,\a_D+\a_2\}\,, \ \ell_2>\ell_3\\\{\a_D+\a_1\,,\a_D+\a_2\,,\a_D+\a_3\}\,,\ \ell_2<\ell_3\end{cases}\,.
\end{split}
\end{align}
After working out the condition \eqref{eq.final}
we find that the reducible points are
\be
\D=\bigg(\bigg\{5+\ell_1+\frac{\ell_2+\ell_3}{2}-\mathbb{Z}_{>0}\bigg\}\bigg\backslash\bigg\{4+\frac{\ell_2+\ell_3}{2}\,,3+\frac{|\ell_2-\ell_3|}{2}
\bigg\}\bigg)\,.
\label{6d}
\ee
In the language of the parametrization $\lambda_1\ge \lambda_2\ge  |\lambda_3|\ge 0$ in Appendix \ref{app.ABD}
this becomes
\begin{align}
\Delta=\bigg\{ \lambda_1+5-\mathbb{Z}_{>0}\bigg\} \setminus \bigg\{\lambda_2+4, |\lambda_3|+3 \bigg\} \;.
\end{align}

\section{Root System for SCAs}\label{app.SCA}
\subsection{\texorpdfstring{3d $\mathcal{N}$ Even}{3d N Even}}\label{sec.3d_even}

The root system for $\mathfrak{g}=D\left(\frac{\scN}{2}, 2\right)$ is given by
\begin{align}
&\Delta_{\bar{0}}=\bigg\{ 
\pm \beta_D , \quad \pm \beta_J, \quad  \pm \beta_D \pm  \beta_J  \;, \quad
\pm \delta_i \pm \delta_j 
\bigg\} \ , \\
&\Delta_{\bar{1}}=\bigg\{
\pm \frac{1}{2}\beta_D \pm \frac{1}{2}\beta_J \pm \delta_i 
\bigg\} \ ,
\end{align}
with $i=1, \ldots, \frac{\scN}{2}$ and the inner product is given by
\begin{align}
(\beta_a, \beta_b)=2\delta_{ab} \;, \quad 
(\delta_i, \beta_a)=0\;, \quad
(\delta_i, \delta_j)=-\delta_{ij}\;, 
\label{3dN_even_inner}
\end{align}
with $a,b=D,J$.
In dictionary ordering $\beta_D> \beta_J > \delta_1> \delta_2>\ldots>\delta_{\frac{\scN}{2}}$, 
we find
\begin{align}
&\Delta_{\fl}=\bigg\{ \pm \beta_J, \quad \pm \delta_i \pm \delta_j \bigg\}\;,\\
&\Delta_{1}^{+}=\overline{\Delta}_{1}^{+}=\bigg\{ \frac{1}{2}\beta_D \pm \frac{1}{2}\beta_J\pm \delta_i  \bigg\}\;, \\
&\Delta_{\fn}\cap \overline{\Delta}_0=\Delta_{\fn}=\bigg\{ \beta_D+\beta_J, \quad \beta_D, \quad \beta_D-\beta_J \bigg\}\;.
\end{align}

\subsection{\texorpdfstring{3d $\mathcal{N}$ Odd}{3d N Odd}}\label{sec.3d_odd}

We have $\mathfrak{g}=B\left(\frac{\scN-1}{2}, 2\right)$.
We have the root system 
\begin{align}
&\Delta_{\bar{0}}=\bigg\{ 
\pm \beta_D , \quad \pm \beta_J, \quad  \pm \beta_D \pm  \beta_J \;,\quad
\pm \delta_i \pm \delta_j, \quad \pm \delta_i
\bigg\} \ , \\
&\Delta_{\bar{1}}=\bigg\{
 \pm \frac{1}{2}\beta_D \pm \frac{1}{2}\beta_J \pm \delta_i, \quad \pm \frac{1}{2}\beta_D \pm \frac{1}{2}\beta_J 
\bigg\} \ ,
\end{align}
with $i=1,\ldots, \frac{\scN-1}{2}$.
For $\scN=1$ the vectors $\delta_i$'s are absent, and correspondingly we disregard those roots containing 
these vectors. The inner product is given by \eqref{3dN_even_inner}, and 
with dictionary ordering $\beta_D> \beta_J > \delta_1> \delta_2>\ldots>\delta_{\frac{\scN-1}{2}}$ we obtain
\begin{align}
&\Delta_{\fl}=\bigg\{ \pm \beta_J \;,  \quad \pm \delta_i \pm \delta_j, \quad \pm \delta_i, \bigg\}\;,\\
&\overline{\Delta}_{1}^{+}=\bigg\{\pm \frac{1}{2}\beta_D \pm \frac{1}{2}\beta_J  \pm \delta_i 
  \bigg\}\;, \\
&\Delta_{1}^{+}\setminus \overline{\Delta}_{1}^{+}=\bigg\{ \pm \frac{1}{2}\beta_D \pm \frac{1}{2}\beta_J 
  \bigg\}\;, \\
&\Delta_{\fn}\cap \overline{\Delta}_0=\Delta_{\fn}=\bigg\{ \beta_D+\beta_J, \quad \beta_D, \quad \beta_D-\beta_J \bigg\}\;.
\end{align}

\subsection{\texorpdfstring{4d $\scN \ge 1$}{4d N>=1}} \label{app.D4N}

For $\mathfrak{g}=\mathfrak{su}(4|\scN)$ ($\mathfrak{g}=\mathfrak{psu}(4|4)$ for $\scN=4$),
it is useful to introduce a basis
\beq
(\varepsilon_i, \varepsilon_j)=\delta_{i,j} \ , \quad
(\varepsilon_i, \delta_b)=0 \ , \quad
(\delta_a, \delta_b)=-\delta_{a, b} \ , 
\eeq
with $i,j=1, \ldots, 4$ and $a, b=1, \ldots, \scN$.
Then we have
\begin{align}
&\Delta_{\bar{0}}=\bigg\{ 
\pm( \varepsilon_i - \varepsilon_j ),   \quad
\pm (\delta_a - \delta_b)
\bigg\} \ , \quad
\Delta_{\bar{1}}=\bigg\{
\pm( \varepsilon_i + \delta_a)
\bigg\} \;,\\
&\Delta_{\fl}=\bigg\{ 
\pm( \varepsilon_1 - \varepsilon_2 ),   \quad
\pm( \varepsilon_3 - \varepsilon_4),   \quad
\pm (\delta_a - \delta_b)
\bigg\}\;.
\end{align}
Following \cite{Dobrev:1985qz,Minwalla:1997ka} we choose 
an ordering 
\begin{align}
\varepsilon_1 > \varepsilon_2 >-\varepsilon_3>-\varepsilon_4>-\delta_1 > \dots >-\delta_{\scN} \;,
\end{align}
We have positive simple roots
\begin{align}
&\bigg\{
\varepsilon_{1}-\varepsilon_{2}  \;, \quad
\varepsilon_{4}-\varepsilon_{3}  \;, \quad
\varepsilon_{2}+\delta_{1}  \;, \quad
-\varepsilon_{4}-\delta_{\scN}\;, \quad
-\delta_{a}+\delta_{a+1} \quad (a=1, \ldots, \scN-1)
\bigg\}\;.
\end{align}
We have 
\begin{align}
&\Delta_{1}^{+}=\overline{\Delta}_{1}^{+}=\bigg\{ \varepsilon_1 - \delta_a  \;, \varepsilon_2 - \delta_a  \;, -\varepsilon_3 + \delta_a  \;, -\varepsilon_4 + \delta_a  \bigg\}\;, \\
&\Delta_{\fn}\cap \overline{\Delta}_0=\Delta_{\fn}=\bigg\{
\varepsilon_1 - \varepsilon_3 ,   \quad
\varepsilon_1 - \varepsilon_4 ,   \quad
\varepsilon_2 - \varepsilon_3 ,   \quad
\varepsilon_2 - \varepsilon_4 
 \bigg\}\;.
\end{align}

The highest weight vector is
\begin{align}
\begin{split}
\l&=-\frac{\D}{2}(\varepsilon_1+\varepsilon_2-\varepsilon_3-\varepsilon_4)+\frac{\ell_1}{2}(\varepsilon_1-\varepsilon_2)-\frac{\ell_2}{2}(\varepsilon_3-\varepsilon_4)\\
&\quad+ \frac{\scN-4}{8 \scN} R \left( \sum_{i=1}^4 \varepsilon_i-  \sum_{a=1}^{\scN} \delta_a \right)
-\sum_{b=1}^{\scN} \left( \lambda_b -\frac{|\lambda|}{\scN} \right)  \delta_b \,,
\end{split}
\end{align}
Here $\ell_1$ and $\ell_2$ are the two angular spins, and take integer values.
The set of integers $\{\lambda_a\}$ define a partition, see Appendix \ref{app.A_root}.
Note the factor with $R$ is absent for the special case of $\scN=4$, where there is a reduction of $\mathfrak{u}(1)$ symmetry, from $\mathfrak{sl}(4|4)$ into $\mathfrak{psl}(4|4)$.

The Weyl vector is given by
\begin{align}
&\r_0=\frac{1}{2} \left(3 \varepsilon_1 +\varepsilon_2 -3\varepsilon_3 -\varepsilon_4 \right)
-\sum_{a=1}^{\scN}\frac{\scN+1-2a}{2}\delta_a\,,\\
&\r_1=\frac{\scN}{2}\left( \varepsilon_1+ \varepsilon_2 - \varepsilon_3- \varepsilon_4 \right)\,,\\
&\r= \left(\frac{3-\scN}{2} \varepsilon_1 +\frac{1-\scN}{2}\varepsilon_2 -\frac{3-\scN}{2}\varepsilon_3 -\frac{1-\scN}{2}\varepsilon_4 \right)-\sum_{a=1}^{\scN} \frac{\scN+1-2a}{2}  \delta_a\,,
\end{align}
We therefore obtain
\begin{align}
\begin{split}
\l+\rho&=\frac{-\Delta+3-\scN}{2} \varepsilon_1 +\frac{-\Delta+1-\scN}{2}\varepsilon_2 -\frac{-\Delta+3-\scN}{2}\varepsilon_3 -\frac{-\Delta+1-\scN}{2}\varepsilon_4 \\
&\quad+\frac{\ell_1}{2}(\varepsilon_1-\varepsilon_2)-\frac{\ell_2}{2}(\varepsilon_3-\varepsilon_4)\\
&\quad+ \frac{\scN-4}{8 \scN} R \left( \sum_{i=1}^4 \varepsilon_i-  \sum_{a=1}^{\scN} \delta_a \right)
-\sum_{a=1}^{\scN} \frac{2\lambda_b -2\frac{|\lambda|}{\scN} +\scN+1-2a}{2}  \delta_a\,.
\label{lambda_rho_N4}
\end{split}
\end{align}

For Step 1, we compute
\begin{align}
\begin{split}
(\lambda+\rho, \varepsilon_1)&=\frac{1}{2} \left(-\Delta+\ell_1+3-\scN+\frac{\scN-4}{4\scN} R\right) \;, \\
(\lambda+\rho, \varepsilon_2)&=\frac{1}{2} \left(-\Delta-\ell_1+1-\scN+\frac{\scN-4}{4\scN} R\right) \;, \\
(\lambda+\rho, \varepsilon_3)&=\frac{1}{2} \left(\Delta-\ell_2-3+\scN+\frac{\scN-4}{4\scN} R\right) \;, \\
(\lambda+\rho, \varepsilon_4)&=\frac{1}{2} \left(\Delta+\ell_2-1+\scN+\frac{\scN-4}{4\scN} R\right) \;, \\
(\lambda+\rho,  \delta_a)&=\frac{1}{2} \left(\frac{\scN-4}{4\scN}R+2\lambda_a -2\frac{|\lambda|}{\scN} +\scN+1-2a \right) \;.
\end{split}
\end{align}

From this we can easily see that 
the module is reducible at \footnote{Compared with  \cite{Minwalla:1997ka} we have an extra minus sign in front of $\ell_2$ in the third line.}
\begin{align}\label{4dN_iso}
\Delta=
\begin{dcases}
 \ell_1+\frac{\scN-4}{2\scN} R   +2 \lambda_a-2\frac{|\lambda|}{\scN} -2a  +4& (\varepsilon_1-\delta_a) \;,\\
 -\ell_1+\frac{\scN-4}{2\scN}R   +2 \lambda_a-2\frac{|\lambda|}{\scN} -2a+2&(\varepsilon_2-\delta_a) \;,\\
  -\ell_2-\frac{\scN-4}{2\scN} R   -2 \lambda_a+2\frac{|\lambda|}{\scN} +2a+2-2\scN& (\varepsilon_3-\delta_a)\;,\\
 \ell_2-\frac{\scN-4}{2\scN} R   -2 \lambda_a+2\frac{|\lambda|}{\scN} +2a-2\scN&(\varepsilon_4-\delta_a) \;.\\
\end{dcases} 
\end{align}

We next come to Step 2$^{\prime}$.

A care is needed in this step since in the expression for $\lambda+\rho$ in \eqref{lambda_rho_N4}
the coefficients of the $\varepsilon_i$ do depend non-trivially on the R-charge $R$.
However, such a R-dependence drops out when we consider irreducibility in this step, since $\sum_i \varepsilon_i  -\sum_a \delta_a$ is orthogonal to all the
roots corresponding to momentum generators. Indeed, we can compute
\begin{align}
\begin{split}
&n_{\varepsilon_1-\varepsilon_3}=  -\Delta +\frac{\ell_1-\ell_2}{2}+3 -\scN \,,\quad
n_{\varepsilon_1-\varepsilon_4}= -\Delta +\frac{\ell_1+\ell_2}{2}+2 -\scN  \,,\\
&n_{\varepsilon_2-\varepsilon_3}=  -\Delta +\frac{-\ell_1-\ell_2}{2}+2 -\scN  \,,\quad
n_{\varepsilon_2-\varepsilon_4}=  -\Delta +\frac{-\ell_1+\ell_2}{2}+1 -\scN  \,.
\end{split}
\end{align}
From this we can work out when  the set $\Psi_{\lambda, {\rm non-iso}}$ is non-empty.
We again learn that only the effect of $\scN$ in the rest of the analysis 
is to shift $\Delta\to \Delta+\scN$. 

The result is then obtained by combining \eqref{4dN_iso}
and \eqref{4d}, where $\Delta$ in the latter is shifted by $\scN$.

\subsection{\texorpdfstring{5d $\scN=1$}{5d N=1}}

For this case, we use the root system for $\mathfrak{f}_4$:
\begin{align}
&\Delta_{\bar{0}}=\bigg\{ 
\pm \beta_D \pm \beta_{J_1},  \quad
\pm \beta_D \pm \beta_{J_2},  \quad
\pm \beta_{J_1} \pm \beta_{J_2},  \quad
\pm \beta_D ,  \quad
\pm \beta_{J_1},  \quad
\pm \beta_{J_2},  \quad
\pm \delta
\bigg\} \ ,\\
&\Delta_{\bar{1}}=\bigg\{
\pm \frac{1}{2}\beta_D \pm \frac{1}{2}\beta_{J_1} \pm \frac{1}{2}\beta_{J_2}\pm \frac{1}{2}\delta 
 \bigg\} \ ,
\end{align}
with the inner product
\beq
(\beta_i, \beta_j)=\delta_{i,j}\;, \quad (\beta_i, \delta)=0 \;, \quad (\delta, \delta)=-3 \;.
\eeq
Under an ordering $\beta_D> \beta_{J_1}>\beta_{J_2}>\delta$,
\begin{align}
&\Delta_{\fl}=\bigg\{ 
\pm \beta_{J_1} \pm \beta_{J_2},  \quad
\pm \beta_{J_1},  \quad
\pm \beta_{J_2},  \quad
\pm \delta
\bigg\}\;,\\
&\Delta_{1}^{+}=\overline{\Delta}_{1}^{+}=\bigg\{  \frac{1}{2}\beta_D \pm \frac{1}{2}\beta_{J_1} \pm \frac{1}{2}\beta_{J_2}  \pm \frac{1}{2}\delta  \bigg\}\;, \\
&\Delta_{\fn}\cap \overline{\Delta}_0=\Delta_{\fn}=\bigg\{ 
\beta_D \pm \beta_{J_1},  \quad
\beta_D \pm \beta_{J_2},  \quad
\beta_D 
 \bigg\}\;.
\end{align}
We have 
\begin{align}
&\lambda=-\Delta \beta_D + \left( \ell_1 +\frac{\ell_2}{2}\right) \beta_{J_1}+\frac{\ell_2}{2} \beta_{J_2} + k\delta\;,\\
&\rho_0=\frac{5}{2}\beta_D+\frac{3}{2}\beta_{J_1}+\frac{1}{2}\beta_{J_2}+\frac{1}{2}\delta \;,\quad
\rho_1=2 \beta_D\;, \\
&\rho=\frac{1}{2}\beta_D+\frac{3}{2}\beta_{J_1}+\frac{1}{2}\beta_{J_2}+\frac{1}{2}\delta\;, \\
&\lambda+\rho=\left(-\Delta+\frac{1}{2}\right)\beta_D+\left(\ell_1+\frac{\ell_2}{2}+\frac{3}{2}\right)\beta_{J_1}+\left(\frac{\ell_2}{2}+\frac{1}{2}\right)\beta_{J_2}+ \left( k+\frac{1}{2} \right)\delta \;.
\end{align}
Step 1 gives
\begin{align}
\Delta=\frac{1}{2}+s_1\left(\ell_1+\frac{\ell_2}{2}+\frac{3}{2}\right)+s_2\left(\frac{\ell_2}{2}+\frac{1}{2}\right)-3\sigma\left( k+\frac{1}{2} \right) \;,
\end{align}
with $s_1, s_2, \sigma=\pm 1$. In Step 2$^{\prime}$ the shift of the value of $\Delta$ is $2$, as originating from the 
coefficient of $\beta_D$ in $\rho_1$.
Hence the module is reducible at these values, as well at at \eqref{5d} with shift of $\Delta$ by $2$.

\subsection{\texorpdfstring{6d $\scN=(0,1)$}{6d N=(0,1)}}

The root system for $\mathfrak{g}=\mathfrak{osp}(8|2)=D(4,1)$ is
\begin{align}
&\Delta_{\bar{0}}=\bigg\{ 
\pm \alpha_i \pm \alpha_j,  \quad  \pm  \beta , \bigg\} \quad (i,j=D,1,2,3)\;, \\
&\Delta_{\bar{1}}=\bigg\{
\frac{s_D}{2}\alpha_D + \frac{s_1}{2}\alpha_1 + \frac{s_2}{2}\alpha_2 + \frac{s_3}{2}\alpha_3  + \frac{\sigma}{2}\beta 
\bigg\} \ ,
\end{align}
where $s_D, s_1, \ldots, s_3, \sigma=\pm 1$ with the constraint $s_D s_1 s_2 s_3=1$,
and we have
\begin{align}
(\alpha_i, \alpha_j)=\delta_{ij}\;, \quad (\beta, \beta)=-2 \;.
\end{align}
Under an ordering $\a_D>\a_1>\a_2>\a_3>\beta$,
\begin{align}
&\Delta_{\fl}=\bigg\{ 
\pm \alpha_1 \pm \alpha_2,  \quad
\pm \alpha_1 \pm \alpha_3,  \quad
\pm \alpha_2 \pm \alpha_3,  \quad  \pm  \beta \bigg\}\;,\\
&\Delta_{1}^{+}=\overline{\Delta}_{1}^{+}=\bigg\{ \frac{1}{2}\alpha_D +\frac{s_1}{2}\alpha_1 + \frac{s_2}{2}\alpha_2 +\frac{s_3}{2}\alpha_3  +\frac{\sigma}{2}\beta  \bigg\}\;, \label{6dN1_iso}\\
&\Delta_{\fn}\cap \overline{\Delta}_0=\Delta_{\fn}=\bigg\{ \alpha_D\pm \alpha_{1}, \quad  \alpha_D\pm \alpha_{2}, \quad \alpha_D\pm \alpha_{3} \bigg\}\;,
\end{align}
with the constraint $s_1 s_2 s_3=1$.
{\small
\begin{align}
&\lambda= -\Delta \alpha_D +\left(\ell_1 +\frac{\ell_2+\ell_3}{2} \right) \alpha_1+\frac{\ell_2+\ell_3}{2}\alpha_2+\frac{-\ell_2+\ell_3}{2} \alpha_3+\frac{k}{2} \beta \;,\\
&\rho=\alpha_D+2 \alpha_1+\alpha_2+\frac{1}{2}\beta\;, \\
&\lambda+\rho=( -\Delta+1) \alpha_D +\left(\ell_1 +\frac{\ell_2+\ell_3}{2}  +2\right)\alpha_1+\left(\frac{\ell_2+\ell_3}{2}  +1\right) \alpha_2+ \frac{-\ell_2+\ell_3}{2}   \alpha_3+\left(\frac{k}{2}+\frac{1}{2}\right) \beta\;.
\end{align}
}
Then
$(\lambda+\rho, \alpha)=0$ for an odd isotropic root $\alpha$ from \eqref{6dN1_iso} gives
\begin{align}
\Delta=1+\left(\ell_1 +\frac{\ell_2+\ell_3}{2}  +2\right) s_1+\left(\frac{\ell_2+\ell_3}{2}  +1\right)  s_2+\frac{-\ell_2+\ell_3}{2}  s_3 +4\sigma \left(\frac{k}{2}+\frac{1}{2}\right) \;.
\end{align}
The module is reducible either at these values or at values \eqref{6d}, with $\Delta$ shifted by $2$.

\subsection{\texorpdfstring{6d $\scN=(0,2)$}{6d N=(0,2)}}

We have $\mathfrak{g}=\mathfrak{osp}(8|4)=D(4,2)$,
which has the root system
\begin{align}
&\Delta_{\bar{0}}=\bigg\{ 
\pm \alpha_i \pm \alpha_j,  \quad
\pm \beta_1 \pm \beta_2, \quad  \pm  \beta_1 , \quad  \pm  \beta_2
\bigg\} \ , \quad (i,j=D,1,2,3)\\
&\Delta_{\bar{1}}=\bigg\{
\frac{s_D}{2}\alpha_D + \frac{s_1}{2}\alpha_1 + \frac{s_2}{2}\alpha_2 + \frac{s_3}{2}\alpha_3  + \frac{\sigma_1}{2}\beta_1 
+\frac{\sigma_2}{2}\beta_2
\bigg\} \;,
\end{align}
where $s_D, s_1, \ldots, s_3, \sigma_1, \sigma_2=\pm 1$ with the constraint $s_D s_1 s_2 s_3=1$,
and with
\begin{align}
(\alpha_i, \alpha_j)=\delta_{ij}\;, \quad (\beta_a, \beta_b)=-\delta_{ij} \;.
\end{align}
Under an ordering $\a_D>\a_1>\a_2>\a_3>\beta_1>\beta_2$,
\begin{align}
&\Delta_{\fl}=\bigg\{ 
\pm \alpha_1 \pm \alpha_2,  \quad
\pm \alpha_1 \pm \alpha_3,  \quad
\pm \alpha_2 \pm \alpha_3,  \quad
\pm \beta_1 \pm \beta_2, \quad  \pm  \beta_1 , \quad  \pm  \beta_2\bigg\}\;,\\
&\Delta_{1}^{+}=\overline{\Delta}_{1}^{+}=\bigg\{ \frac{1}{2}\alpha_D + \frac{s_1}{2}\alpha_1 + \frac{s_2}{2}\alpha_2 +\frac{s_3}{2}\alpha_3  + \frac{\sigma_1}{2}\beta_ 1+ \frac{\sigma_2}{2}\beta_ 2  \bigg\}\;,  \label{6dN2_iso}\\
&\Delta_{\fn}\cap \overline{\Delta}_0=\Delta_{\fn}=\bigg\{ \alpha_D\pm \alpha_{1}, \quad  \alpha_D\pm \alpha_{2}, \quad \alpha_D\pm \alpha_{3} \bigg\}\;.
\end{align}
with the constraint $s_1 s_2 s_3=1$.
We have
\begin{align}
&\lambda= -\Delta \alpha_D +\left(\ell_1 +\frac{\ell_2+\ell_3}{2} \right) \alpha_1+\frac{\ell_2+\ell_3}{2}  \alpha_2+\frac{-\ell_2+\ell_3}{2} \alpha_3+k_1 \beta_1+k_2 \beta_2\;, \\
&\rho=-\alpha_D+2 \alpha_1+\alpha_2+\frac{3}{2}\beta_1+\frac{1}{2}\beta_2\;,\\
&\lambda+\rho=( -\Delta-1) \alpha_D +\left(\ell_1 +\frac{\ell_2+\ell_3}{2}+2\right) \alpha_1+ \left(\frac{\ell_2+\ell_3}{2}+1 \right) \alpha_2+\frac{-\ell_2+\ell_3}{2}  \alpha_3 \nonumber \\
&\qquad +\left(k_1 +\frac{3}{2}\right) \beta_1+\left(k_2 +\frac{1}{2}\right) \beta_2\;.
\end{align}
Then
$(\lambda+\rho, \alpha)=0$ for an odd isotropic root $\alpha$ from \eqref{6dN2_iso} gives
\begin{align}
&\Delta=-1+\left(\ell_1 +\frac{\ell_2+\ell_3}{2}+2\right) s_1+ \left(\frac{\ell_2+\ell_3}{2}+1 \right)  s_2+\frac{-\ell_2+\ell_3}{2} s_3 \nonumber \\& \qquad +2\sigma_1 \left(k_1 +\frac{3}{2}\right) +2\sigma_2 \left(k_2 +\frac{1}{2}\right)\;.
\end{align}
The module is reducible either at these values or at \eqref{6d} with
$\Delta$ shifted by minus $4$.


\bibliographystyle{nb}
\bibliography{bootstrap_unitary}


\end{document}